\documentclass[%
reprint,
superscriptaddress,
amsmath,amssymb,
aps,
]{revtex4-2}

\usepackage[dvipsnames]{xcolor}
\usepackage[
	colorlinks=true,
	citecolor=Red,
	linkcolor=Blue,
	filecolor=magenta,
	urlcolor=Blue,
]{hyperref}

\usepackage{physics}
\usepackage{graphicx}
\usepackage{comment}
\usepackage{xfrac}

\usepackage{lipsum}

\DeclareMathAlphabet\mathbfcal{OMS}{cmsy}{b}{n}
\newcommand{\qqoute}[1]{``#1''}

\newcommand{\iu}{{i}\mkern1mu}
\newcommand{\eu}{\mathrm{e}\mkern1mu}

\definecolor{pnasbluetext}{RGB}{0,101,165} %
\definecolor{pnasblueback}{RGB}{205,217,235} %
\newcommand{\ten}[1]{\boldsymbol{\vb{\hat{#1}}}}

\newcommand{\ITMO}{School of Physics and Engineering, ITMO University, St. Petersburg 197101, Russia} 

\newcommand{\ANU}{Nonlinear Physics Centre, Research School of Physics, The Australian National University, Canberra, ACT 2601,
Australia}

\begin{document}

\title{Acoustic Lateral Recoil Force and Stable Lift of Anisotropic Particles}

\author{Mikhail Smagin}

\affiliation{Qingdao Innovation and Development Center, Harbin Engineering
University, Qingdao 266000 Shandong, China}
\affiliation{\ITMO}

\author{Ivan Toftul}

\affiliation{\ITMO}
\affiliation{\ANU}

\author{Konstantin Y. Bliokh}
\affiliation{Donostia International Physics Center (DIPC), Donostia-San Sebasti\'{a}n 20018, Spain}
\affiliation{Theoretical Quantum Physics Laboratory, Cluster for Pioneering Research, RIKEN, Wako-shi, Saitama 351-0198, Japan}
\affiliation{Centre of Excellence ENSEMBLE3 Sp. z o.o., 01-919 Warsaw, Poland}

\author{Mihail Petrov}
\affiliation{\ITMO}
\affiliation{Qingdao Innovation and Development Center, Harbin Engineering
University, Qingdao 266000 Shandong, China}
\email{m.petrov@metalab.ifmo.ru
}

\begin{abstract}
Acoustic forces and torques are of immense importance for manipulation of particles, in particular in biomedical applications. While such forces and torques are well understood for small spherical particles with lowest-order monopole and dipole responses, the higher-order effects for larger anisotropic particles have not been properly investigated. Here we examine the acoustic force and torque on an anisotropic (spheroid) particle and reveal two novel phenomena. First, we describe the {\it lateral recoil force}, orthogonal to the direction of the incident wave and determined by the tilted orientation of the particle. Second, we find conditions for the {\it stable acoustic lift}, where the balanced torque and force produce a stable lateral drift of the tilted particle. We argue that these phenomena can bring about new functionalities in acoustic manipulation and sorting of anisotropic particles including biological objects such as blood cells.

\end{abstract}

\maketitle

\section*{Introduction} 
\label{sec:intro}
\vspace{-5pt}
Manipulation of objects by means of acoustic waves recently showed significant progress, demonstrating great potential for biomedical applications, sorting of cells, acoustic tweezers, controlled levitation, and volumetric displays~\cite{life-sciences-review, tweezer-review, tweezer-review-2,Fan2022Sep,Rufo2022Apr,Wu2019Jun,Olofsson2020,Mura2013Nov, Maresca2018Jun, Wu2023Feb,Hirayama2019Nov, Fushimi2019Aug, Hossein2023Aug}. While acoustic tweezers are typically used to trap isotropic spherical particles~\cite{Wixforth2004Aug, Shi2009Oct, Guo2016Feb, Hirayama2019Nov}, many systems involve particles with complex shapes~\cite{Wu2017Oct, Ahmed2016Mar, Li2016Nov, Dow2018Mar, Urbansky2017Dec,Aghakhani2022Mar,Tian2020Sep} including biological objects~\cite{Jooss2022Mar}. Therefore, theoretical~\cite{silva, powell, Sepehrirahnama2022Oct, Sepehrirahnama2022May, Tang2023Jun} and experimental~\cite{Tang2023Jun, Ren2019Oct} studies of acoustic field interactions with {\it anisotropic} particles is currently in high demand. 

Acoustic manipulation of complex-shape objects can be challenging, yet it opens ways for novel functionalities and additional degrees of control. Still, existing theoretical approaches to acoustic forces lack rigorous analysis of nontrivial multipole interference in the scattered field and related acoustomechanical effects. Furthermore, very few studies have properly examined acoustic {\it torques}.

In this work, we examine acoustic forces and torques on anisotropic particles, such as spheroid (rotational ellipsoid) particles. 
We employ rigorous multipole analysis which has shown its efficiency in nanophotonics~\cite{Liu2020May, Kislov2021Sep, Gladyshev2020} and acoustics~\cite{Wu2021Aug, baresch, Tsimokha2022Apr}.

\begin{figure}[h]
\includegraphics[width=1.0\linewidth]{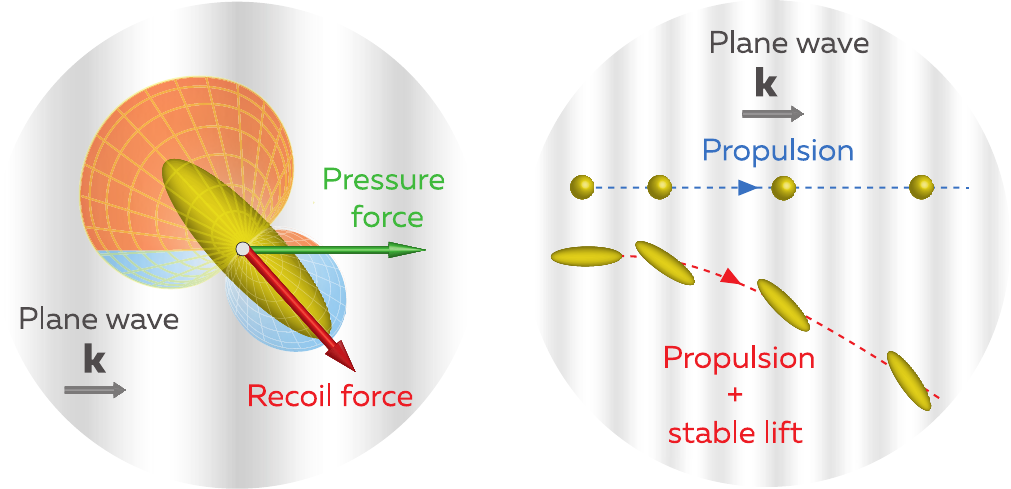}
\caption{\label{fig1} Schematics of the problem. (a) An anisotropic particle in an incident acoustic plane-wave field exhibits an asymmetric scattering response resulting in the longitudinal and lateral acoustic force components. (b) Stable acoustic lift appears when the particle drifts in the lateral direction without rotation (acoustic torque is balanced).}
\vspace{-7pt}
\end{figure}

Importantly, we reveal a {\it lateral recoil force} directed orthogonally to the incident-wave direction and originating from the interference of the monopole and dipole contributions to the scattered field, Fig.~\ref{fig1}. This effect is an acoustic counterpart of the optical recoil force originating from the interference of electric and magnetic dipole contributions~\cite{chaumet,nieto-optics-express,optical-pulling-ct-chan,Bliokh2014Mar,Bekshaev2015Mar,Antognozzi2016Aug,Yevick2016Apr,Hayat2015Oct}.
Moreover, we find that the lateral force can be accompanied by a balanced torque, thereby realizing {\it stable acoustic lift}, Fig.~\ref{fig1}. In this case, the tilted particle moves perpendicular to the incident wave without rotations, akin to the optical lift~\cite{optical-lift-nature, optical-lift-optica}.

The paper is organized as follows. 
First, we derive the recoil (self-action) corrections to the acoustic force and torque in the general form involving the wave-induced monopole and dipole moments of the particle.
Second, we investigate the lateral-force effect for small spheroid particles. 
Third, we analyse the accompanying torque and find conditions for the stable acoustic lift (involving higher-order corrections).  
Finally, we briefly discuss potential application of this effect for sorting of blood cells or other anisotropic particles and summarize our findings.

\section*{Results}
\section{Acoustic Force and Torque with recoil contributions}

\label{sec:recoil_force}

The most universal method to calculate the radiation force and torque from a linear acoustic wavefield on an arbitrary scattering body is the integration of the acoustic momentum and angular momentum flux tensors over a surface $S$ enclosing the scatterer \cite{Westervelt1951, bruus-7}.  
In this manner, the time-averaged acoustic force and torque in a monochromatic sound wavefield read \cite{PRL,bruus-2,bruus-7}:
\begin{equation}
\label{force_tensor}
{\vb{F}} = - \oint\limits_{S}  \ten{\mathcal{T}} \cdot \dd \vb{S}, \qquad 
{\vb{T}} = - \oint\limits_{S} \ten{\mathcal{M}} \cdot \dd \vb{S},
\end{equation}
where $\ten{\mathcal{T}} = \frac{1}{4} \ten{I} \left( \beta \abs{p_{\text{tot}}}^2 - \rho \abs{\vb{v}_{\text{tot}}}^2 \right) + \frac{1}{2} \rho \Re\left( \vb{v}_{\text{tot}}^* \otimes \vb{v}_{\text{tot}} \right)$  is the cycle-averaged momentum flux density, $\ten{\mathcal{M}} = \vb{r} \times \ten{\mathcal{T}}$ is the corresponding angular momentum flux density; $p_{\text{tot}}(\vb r), \vb{v}_{\text{tot}} (\vb r)$ are the complex amplitudes of the total (i.e., incident + scattered) pressure and velocity fields, 
$\vb{v}^* \otimes \vb{v}$ should be understood as the outer product of two vectors, $\{v^*_i v_j\}$,
and $\hat{I}_{ij} = \delta_{ij}$
is the unit dyadic tensor. 
The real time-dependent pressure and velocity fields are given by $\bar{p}(\vb{r}, t) = \Re [p(\vb{r}) \eu^{- \iu \omega t}]$ and $\bar{{\bf v}}(\vb{r}, t) = \Re [{\bf v}(\vb{r}) \eu^{- \iu \omega t}]$, where $\omega$ is the wave frequency.

\begin{figure}[t]
\includegraphics[width=1.0\linewidth]{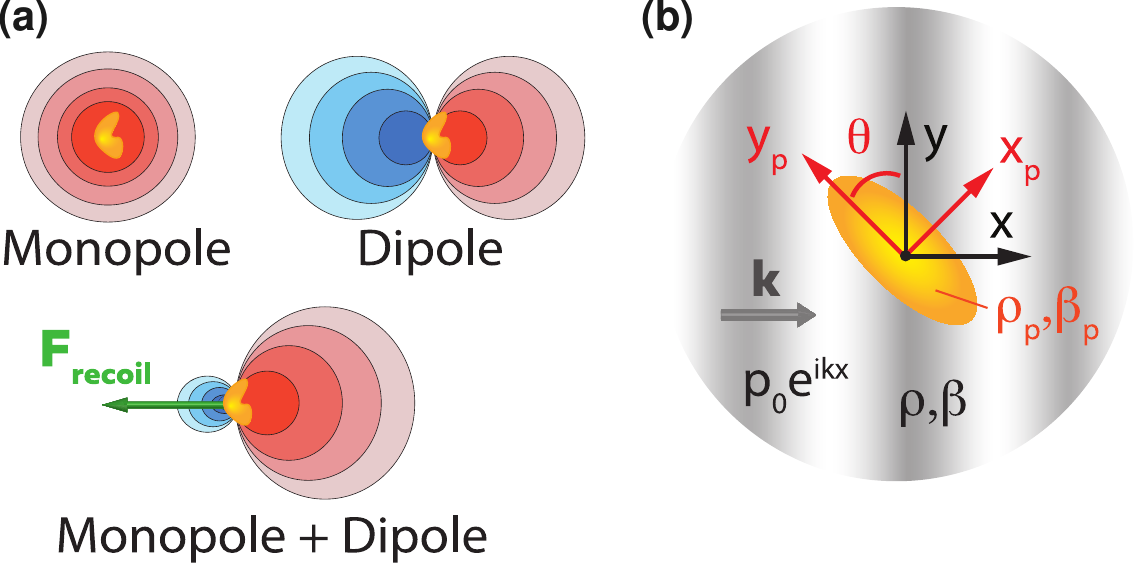}
\caption{(a) Acoustic Kerker effect and its mechanical manifestation. Directional scattering gives rise to a recoil force in the direction opposite to the maximum of the scattered energy flow. (b) The geometry of the problem with a tilted spheroid particle in an incident plane-wave field.}
\label{fig:kerker}
\end{figure}

Expressions \eqref{force_tensor} are valid for particles of any size and shape, and they require solution of the full scattering problem to obtain the total wave field around the particle. In most cases, this can only be done via numerical methods. However, it is often useful to solve this problem in some approximation to obtain simplified analytical expressions for the radiation force and torque in terms of the incident field and properties of the particle. 

We  decompose the total field outside the particle as the sum of incident and scattered fields: $(p_{\text{tot}}, \vb{v}_{\text{tot}}) = (p_{\text{inc}}, \vb{v}_{\text{inc}}) + (p_{\text{sc}}, \vb{v}_{\text{sc}})$.
Substituting this decomposition into Eq.~\eqref{force_tensor} yields three terms in the decomposition of the momentum-flux tensor: $\mathbf{F} = - \oint_{S} \left( \ten{\mathcal{T}}_{\text{inc}} + \ten{\mathcal{T}}_{\text{mix}} + \ten{\mathcal{T}}_{\mathrm{recoil}} \right)\cdot \dd \vb{S}$. 
Here $\ten{\mathcal{T}}_{\mathrm{inc}}$ includes only the incident field and is integrated to zero because there is no change of the momentum due to the free-space propagation; $\ten{\mathcal{T}}_{\mathrm{mix}}$ includes terms which involve both the incident and scattered fields; and $\ten{\mathcal{T}}_{\text{recoil}}$ contains only the scattered fields and, thus, describes {\it self-interaction}.

In general, the scattered field can be decomposed into a sum of multipole (monopole, dipole, quadrupole, etc.) contributions \cite{williams, blackstock}. For particles small compared to the wavelength, $ka \ll 1$ ($k$ is the wavenumber, $a$ is the chracteristic size of the particle), it is sufficient to consider the \textit{monopole} $M$ and \textit{dipole} $\vb{D}$ moments which are proportional to the particle volume, $\propto (ka)^3$ \cite{bruus-7,PRL,Silva2014Nov}. Higher-order multipole contributions scale as $\propto (ka)^{2n+1}$, where $n$ is the multipole order, and can be neglected~\cite[see Table~I in SM]{PRL}.

Here we derive the force and torque on a small acoustic particle which exhibits only the monopole and dipole moments. 
The mixed term $\ten{\mathcal{T}}_{\mathrm{mix}}$ in the force can be integrated in the near-field zone using the gradient expansion of the incident field near the particle center. The resulting expression is known in the literature as a sum of pure monopole and dipole contributions \cite{bruus-7, powell, PRL}.
In contrast, the \textit{self-action} term $\ten{\mathcal{T}}_{\mathrm{recoil}}$ originates from the interference of the monopole and dipole contributions and is usually neglected \cite{bruus-7, powell, Muller2013Aug, Nama2015Jun, Hsu2020Sep} due to the fact that it has characteristic size dependence $\propto (ka)^6$ in the Rayleigh-particle approximation. 
Calculating the radiation force from the mixed and recoil parts of the momentum-flux tensor, we obtain (see Appendix~\ref{app:force_derivation}):
\begin{equation}
{\vb{F} = \underbrace{\frac{1}{2} \Re \left[ M^* \grad p + \rho \vb{D}^* \cdot (\grad) \vb{v} \right]}_{ \vb{F}_{\text{M}} + \vb{F}_{\text{D}}}  - \underbrace{\frac{ k^4}{12 \pi} \sqrt{\frac{\rho}{\beta}} \Re \left[ M^* \vb{D} \right]}_{\vb{F}_{\text{recoil}}}.}
\label{eq:full_force}
\end{equation}
Here we omit the index \qqoute{inc} for brevity: $p \equiv p_{\text{inc}}$ and $\vb{v} \equiv \vb{v}_{\text{inc}}$. Note that we define the monopole moment in a slightly different way, which is related to the conventional monopole definition as $Q = \dd M / \dd t = - \iu  \omega M$~\cite{PRL,williams}.

The new recoil term in Eq.~(\ref{eq:full_force}) is an acoustic counterpart of the similar term in the optical radiation force, which is produced by the interference of the electric and magnetic dipole moments~\cite{chaumet,nieto-optics-express,optical-pulling-ct-chan,Bliokh2014Mar,Bekshaev2015Mar,Antognozzi2016Aug,Yevick2016Apr,Hayat2015Oct} and was not presented in such form in the previous works \cite{Lima_2020}.
The interference of the monopole and dipole radiation can result in strongly directional scattering of the field giving rise to the acoustic recoil force. For example, Fig.~\ref{fig:kerker} shows the in-phase interference between the monopole and dipole contributions resulting in the zero backscattering. This effect is analogous to the Kerker effect in optics~\cite{kerker}, and it was recently explored in acoustics~\cite{Wei2020Aug, Wu2021Aug, Long2020Jun}. 
In this case, the force  $\mathbf{F}_{\mathrm{recoil}}$ will be directed oppositely to the direction of the $k$-vector and have purely {\it pulling} nature~\cite{optical-pulling-ct-chan,Yevick2016Apr}. 
In the approximation under consideration, the total force will be still pushing the particle along the $k$-vector due to the main $\vb{F}_{\text{M}}$ and $\vb{F}_{\text{D}}$ terms. However, proper engineering  of the incident and scattered fields beyond the considered limit can lead to the acoustic pulling~\cite{acoustic-pulling, positive-pressure, acoustic-pulling-active, Fan2021Jul}.

For the case considered in this paper (lossless particle in a plane-wave field), the recoil force is of the same order as the conventional monopole and dipole terms, $\propto (ka)^6$. However, it should be noted that for lossy particles in an inhomogeneous incident field, the higher-order multipole contributions, neglected in the monopole-dipole approximation, can compete with the recoil force (see Appendix~\ref{app:magnitude}). For example, the quadrupole force contribution can have the $(ka)^5$ dependence, since the quadrupole polarizability is connected to the second Mie scattering coefficient $a_2$ with $\Re (a_2) \propto   (ka)^{10}$ and $\Im (a_2) \propto  (ka)^{5}$, see~\cite[see Table 1 in Appendix~\ref{app:magnitude}]{PRL} and \cite{Silva2014Nov}.
In optics, the higher-order quadrupole force and torque, as well as the connection between the Mie coefficients and polarizabilities, can be found in Refs.~\cite{Chen2011Sep, Jiang2016Apr, Evlyukhin2012Jun, LeRu2013Jan}

Using the same approach, but now integrating the angular momentum flux $\ten{\mathcal{M}}$ in Eq.~\eqref{force_tensor}, we derive expression for the acoustic torque in the monopole-dipole approximation with the self-action taken into account:  
\begin{equation}
{\vb{T} = \underbrace{\frac{\rho}{2} \Re \left( \vb{D}^* \cp \vb{v} \right)}_{\vb{T}_{\text{D}}} - \underbrace{\frac{\rho k^3}{24 \pi} \Im \left( \vb{D}^* \cp \vb{D} \right)}_{\vb{T}_{\text{recoil}}}.}
\label{eq:torque_corr}
\end{equation}
The monopole radiation does not contribute to the torque since it carries no angular momentum. The main dipole term $\vb{T}_{\text{D}}$ has been known before \cite{PRL}, while the recoil term $\vb{T}_{\text{recoil}}$ is novel. This term also has an optical counterpart~\cite{Nieto-Vesperinas2015Jul,Nieto-Vesperinas2015Oct}, and it is crucial for the angular momentum conservation. 
In particular, the radiation torque on an isotropic lossless particle must vanish, and this is ensured by the balance of pure dipole and recoil contributions~\cite{PRL,Zhang2014Dec,Zhang2011Apr,Maidanik1958Jul,Silva2014Nov}. Equation \eqref{eq:full_force} and~\eqref{eq:torque_corr} are the key general results of this work. %

\section{Acoustic Force on a Small Spheroid Particle}
\label{subsec:small_ellipsoid}
\subsection{General equations}

We now analyze acoustic forces acting on the simplest example of anisotropic particle: a spheroid (rotational ellipsoid) with two independent semi-axes $b>a=c$, Fig.~\ref{fig:kerker}(b). The case of a rigid spheroid was previously treated using spheroidal wave expansion \cite{Leao-Neto2020Apr,Lima_2020} and is equivalent to our approach. The material parameters of the particles are the density $\rho_\text{p}$, compressibility $\beta_\text{p}$, and the speed of sound $c_{\text{p}} = 1/\sqrt{\rho_\text{p} \beta_\text{p}}$, whereas the  corresponding parameters of the surrounding medium are $\rho$, $\beta$, $c = 1/\sqrt{\rho \beta}$. We assume the incident field to be an $x$-propagating plane wave: 
\begin{equation}
\label{plane wave}
p = p_0 \eu^{\iu kx}\,\quad
\mathbf{v} = v_0\, \bar{\bf x}\, \eu^{\iu kx}\,.
\end{equation}
Here $p_0$ and $v_0 = p_0 \sqrt{{\beta}/{\rho}}$ are the pressure and velocity field amplitudes, whereas the overbar denotes the unit vector of the corresponding axis. 

For any small anisotropic particle, one can express the monopole and dipole moments via the incident-wave fields as~\cite{PRL,Quan2018Jun}
\begin{equation}
\label{monopole_dipole_definition}
{M =  \beta \alpha_{\text{M}} p, \qquad 
\vb{D} = \ten{\alpha}_{\text{D}} \vb{v},}
\end{equation}
where $\alpha_{\text{M}}$ is a scalar monopole polarizability, and $\hat{\boldsymbol{\alpha}}_{\text{D}}$ is the dipole polarizability dyadic tensor. 
For an isotropic spherical particle, $\ten{\alpha}_{\text{D}} = \alpha_{\text{D}} \ten{I}$. 
For a spheroid particle whose longer semi-axis $a$ is aligned with the $y$-axis, the dipole polarizability is diagonal:  $\ten{\alpha}_{\text{D}} = \mathrm{diag} \{\alpha^{\text{D}}_s, \alpha^{\text{D}}_l, \alpha^{\text{D}}_s \}$. Here $\alpha^{\text{D}}_{l}$ and $\alpha^{\text{D}}_{s}$ correspond to the dipolar responses for the velocity field aligned with the longer and shorter semi-axes of the spheroid. 

Without loss of generality, we can assume that the particle is rotated by the angle $\theta$ with respect to the incident wave in the $(x,y)$ plane, as shown in Fig.~\ref{fig:kerker}(b). Then, the dipole polarizability tensor is no longer diagonal and it becomes 
\begin{equation}
{\ten{\alpha}_{\text{D}} = \begin{pmatrix}
\alpha^{\text{D}}_s \cos^2\!\theta + \alpha^{\text{D}}_l \sin^2\!\theta & (\alpha^{\text{D}}_s - \alpha^{\text{D}}_l ) \sin\theta \cos\theta & 0 \\
(\alpha^{\text{D}}_s - \alpha^{\text{D}}_l ) \sin\theta \cos\theta & \alpha^{\text{D}}_l \cos^2\!\theta + \alpha^{\text{D}}_s \sin^2\!\theta & 0 \\
0 & 0 & \alpha^{\text{D}}_s
\end{pmatrix}\!.}
\label{polarizability-matrix}
\end{equation}
Substituting this into Eqs.~(\ref{plane wave}) and (\ref{monopole_dipole_definition}) yields the dipole moments longitudinal and transverse with respect to the incident velocity field:
\begin{align}
{\bf D}_\parallel = v_0\! \left[\alpha^{\text{D}}_l \sin^2\!\theta + \alpha^{\text{D}}_s \cos^2 \!\theta \right] \bar{\bf x}, \nonumber \\ 
{\bf D}_\perp = v_0\! \left(\alpha^{\text{D}}_s - \alpha^{\text{D}}_l \right) \sin\theta \cos\theta\, \bar{\bf y}.
\label{dipole}
\end{align}
\begin{figure*}[t!]
\includegraphics[width=1.0\linewidth]{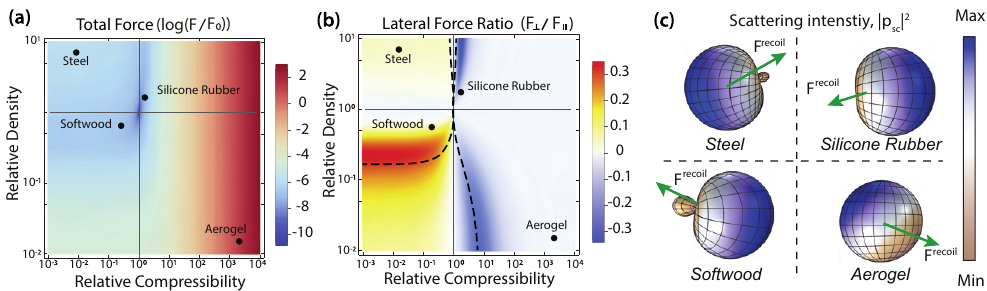}
\caption{\label{fig:parameter_map}Acoustic force on a Rayleigh prolate spheroid particle with $ka = 0.02$, $kb = 0.1$ oriented at the angle $\theta = \pi/4$ with respect to the incident plane wave vs. relative density and compressibility of the particle. The black dots indicate examples of specific particle materials, assuming water as the host material with $c = 1500$~m/s, $\rho = 997~\mathrm{kg/m}^3$. (a) The absolute value of total acoustic force $F = |{\bf F}|$ normalized by the pressure force on a spheroid particle $F_0$ (see explanation in the text). (b) The ratio of the lateral and longitudinal force components, $F_\perp / F_\parallel$. Since the longitudinal force is always positive, there are areas of negative and positive lateral force. The dashed curves show an approximate analytical solution for the extremum values of the particles' parameters, Eq.~\eqref{eq:beta_cr}. (c) Examples of the radiation diagrams for the scattered acoustic intensity for particles made of specific materials in the water environment. The direction of the recoil force, opposite to the scattered momentum, depends on the relative phase of the excited monopole and dipole moments.}
\end{figure*}

Substituting Eqs.~(\ref{plane wave})--(\ref{dipole}) into  Eq.~\eqref{eq:full_force}, we find the monopole, dipole, and recoil parts of the radiation force on the spheroid:
\begin{widetext}
\begin{align}
\mathbf{F}_\mathrm{M} & = \frac{1}{2} k \beta |p_0|^2 \, \mathrm{Im}\! \left( \alpha_{\text{M}} \right) \bar{\bf x}\,, \qquad
\mathbf{F}_\mathrm{D} = \frac{1}{2} k \beta \abs{p_0}^2 
\left[\Im\!\left( \alpha^{\text{D}}_l \right) \sin^2\!\theta  + \Im\!\left( \alpha^{\text{D}}_s \right) \cos^2\!\theta  \right] \bar{\bf x}\,,
\label{eq:dipole_force_polarizability}
\\
\mathbf{F}_\mathrm{recoil} & = - \frac{\beta k^4 |p_0|^2}{12 \pi} \, \Re\!\left[\alpha_{\text{M}}^* (\alpha^{\text{D}}_l \sin^2\!\theta + \alpha^{\text{D}}_s \cos^2\!\theta) \right] \bar{\bf x}
- \frac{\beta k^4 |p_0|^2}{12 \pi} \sin\theta \cos\theta \, \Re\!\left[ \alpha_{\text{M}}^* (\alpha^{\text{D}}_s  - \alpha^{\text{D}}_l) \right] \bar{\bf y}\,.
\label{eq:recoil_force_polarizability}
\end{align}
\end{widetext}
These equations show that the monopole and dipole forces are directed along the incident plane wave, while the recoil force generally has a nonzero {\it lateral force} component. 

In the Appendix~\ref{app:polarizability}, we derive the monopole and dipole polarizabilities of a small ellipsoid particle:
\begin{align}
\label{eq:polariz}
({\alpha_{\text{M}}})^{-1} = ({\alpha_{\text{M}}^{\text{st}}})^{-1} - \dfrac{\iu k^3}{ 4\pi},~~
\left(\ten{\alpha}_{\text{D}}\right)^{-1} = \left(\ten \alpha_{\text{D}}^{_{\text{st}}}\right)^{-1} - \dfrac{\iu k^3}{12\pi} \ten{I}.
\end{align}
Here 
\begin{equation}
\label{eq:static_monopole_dipole}
{\alpha_{\text{M}}^{\text{st}} = V_{\text{e}} \ (\bar{\beta} - 1), \quad
\alpha^{\text{st}}_{\text{D},i}  = V_{\text{e}} \frac{\bar{\rho} - 1}{\bar{\rho} + L_i \left[1 - \bar{\rho} \right]}}
\end{equation}
are the static polarizabilities, where $\bar{\beta} = \beta_{\text{p}}/\beta$ and $\bar{\rho} = \rho_{\text{p}}/\rho$ are the relative compressibility and density of the particle, $V_{\text{e}} = 4 \pi \, a b c/3$ is the ellipsoid volume,  $L_{\{1,2,3\}} = (abc/2) \int\limits_0^\infty \left[ \left(\{c,a,b\}^2 + q \right) \!f(q) \right]^{-1} \dd q$, $f(q)=\sqrt{(q+a^2)(q+b^2)(q+c^2)}$, and $\alpha^{\text{st}}_{\text{D},i}$ are the diagonal components of $\ten{\alpha}_{\text{D}}^{_{\text{st}}}$ for the coordinate axes aligned with the principal axes of the ellipsoid. For spherical particle, the geometric factors $L_1 = L_2 = L_3 = 1/3$, and Eqs.~\eqref{eq:static_monopole_dipole} agree with previously known results~\cite{baresch, PRL}.

\begin{figure}[h]
\includegraphics[width = 1.0\linewidth]{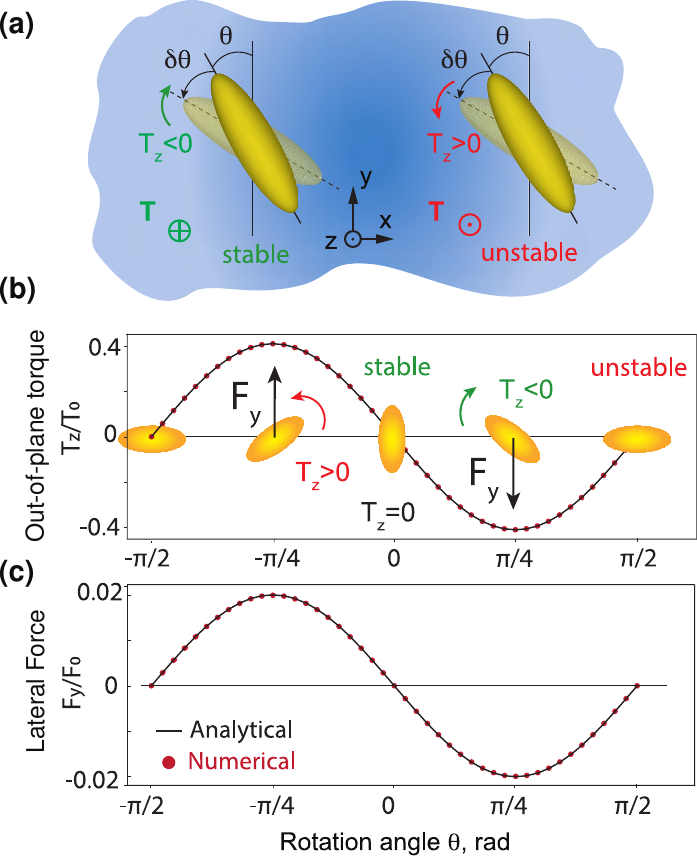}
\caption{\label{fig:Rayleigh_acoustomechanics} (a) Schematics of rotationally stable and unstable particle configurations. The stable configuration is achieved when the torque vanishes and small rotational perturbations induce restoring torques. (b,c) Acoustic torque and lateral force on a prolate Rayleigh spheroid particle vs. the orientation angle $\theta$. The only stable point $\theta = 0$ is trivial, and the lateral force vanishes there.}
\end{figure}
The second terms in the right-hand sides of Eqs.~\eqref{eq:polariz}  are the radiation corrections due to rescattering of the field, which are well known in optics and are always present due to the optical theorem requirements \cite{bohren-huffman, Draine1988Oct, Simpson2010Sep, Belov2003Sep, correction-2, correction-3}. These corrections describe non-zero radiation force on small lossless particles $\propto \Im(\alpha_{\text{M},\text{D}})$~\cite{Westervelt1957Jan}. In acoustics, the radiation corrections to the polarizabilities were previously derived for spheres via Taylor expansion of the Mie coefficients in the Generalized Lorenz-Mie Theory~\cite{baresch,Silva2014Nov}. 

The acoustic dipole polarizability tensor for an ellipsoid particle, Eqs.~\eqref{polarizability-matrix}, \eqref{eq:polariz}, \eqref{eq:static_monopole_dipole}, and the corresponding radiation forces \eqref{eq:dipole_force_polarizability}, \eqref{eq:recoil_force_polarizability} are the main results of this section.

\subsection{Lateral force}

The most interesting outcome of the expressions for the acoustic force on a small anisotropic particle is the lateral recoil force component $F_\perp \equiv F_y$ in Eq.~\eqref{eq:recoil_force_polarizability}. Note that the Willis' coupling correction to acoustic \textit{gradient}~\cite{PRL, Du2017Dec} force considered in \cite{powell, Sepehrirahnama2022Oct, Sepehrirahnama2022May} does not include the recoil contribution which leads to the emergence of lateral force even in fields with no lateral energy gradient such as plane waves. Here we analyse this qualitatively novel force contribution. 
First, since $F_\perp (\theta) \propto \sin 2\theta$, the lateral force is maximum at the orientation angle $\theta=\pi/4$.

Next, Figure~\ref{fig:parameter_map}(a) shows the absolute value of the total acoustic force, $F = |{\bf F}|$, Eqs.~\eqref{eq:full_force}, \eqref{eq:dipole_force_polarizability}, and \eqref{eq:recoil_force_polarizability}, as dependent on the relative density $\bar{\rho}$ and compressibility $\bar{\beta}$ for $\theta=\pi/4$. The plot is divided into four quadrants $\bar{\rho}\lessgtr 1$, $\bar{\beta}\lessgtr 1$ corresponding to different types of acoustic materials. The total force in this and all subsequent figures is normalized by $F_0 = \sigma_{\text{geom}} \beta \abs{p_0}^2 /2$, which is the plane-wave radiation pressure force acting on a particle with the extinction cross-section equal to its geometrical cross-section $\sigma_{\text{geom}} = \pi a b$. 

The effectiveness of the lateral force can be characterized by its ratio to the longitudinal force, $F_\perp/F_\parallel \equiv F_y/F_x$. The dependence of this ratio on $\bar{\rho}$ and $\bar{\beta}$ is shown in Fig.~\ref{fig:parameter_map}(b). Note that in our scenario this parameter map is fully independent of the dimensionless parameter $ka$ (see Appendix \ref{app:magnitude}). While the longitudinal force is always positive, the transverse force changes its sign across the line $\bar{\beta}=1$. 
Moreover, one can see that the effectiveness $F_\perp/F_\parallel$ has pronounced extrema and can be considerably enhanced for specific parameter values [the red and blue zones in Fig.~\ref{fig:parameter_map}(b)], up to one third of the longitudinal force. 

Analyzing  Eqs.~\eqref{eq:dipole_force_polarizability} and \eqref{eq:recoil_force_polarizability} in the Rayleigh limit $ka\ll 1$ and for $\theta \ll 1$, we find the condition for the maximum of $\abs{F_{\perp} / F_{\parallel}}$, which yields a compact formula
\begin{equation}\label{eq:beta_cr}
    \bar{\beta}^{\text{cr}}_{1,2} = 1 \pm   \frac{\abs{\bar{\rho}^{\text{cr}} -1}}{{\sqrt{3}}\abs{L_s (1 - \bar{\rho}^{\text{cr}}) + \bar{\rho}^{\text{cr}}}}.
\end{equation}
These curves, indicating the maximum effectiveness of the lateral force are shown in Fig.~\ref{fig:parameter_map}(b). Even though the color map is for $\theta = \pi/4$, the curves provide a reasonable approximation for the areas of maximum lateral force ratio.

Finally, to explain the physical mechanism underlying the effective lateral force, in Fig.~\ref{fig:parameter_map}(c) we plot scattered-field diagrams and the recoil force direction for spheroid particles made of different materials, which are also marked in Figs.~\ref{fig:parameter_map}(a,b). One can see that these diagrams differ from each other because of the difference in the relative phases and magnitudes of the excited dipole and monopole moments.  
The most effective lateral force appears for the parameters corresponding to the negative longitudinal recoil force $F_{{\rm recoil},x} < 0$, suppressing the main radiation force $F_{{\rm M},x}+F_{{\rm D},x}>0$ (the acoustic Kerker effect), and high-magnitude lateral force $F_{{\rm recoil},y}$ (see the particleboard softwood case). 

\section{Stable Acoustic Lift}
\label{sec:stable_lift}

The acoustomechanical behaviour of a particle is determined by both the acoustic radiation force and torque. 
In particular, to produce a stable lateral drift (lift) of the particle, there should be a nonzero lateral force (which requires a tilted orientation of the spheroid particle), as well as vanishing torque to avoid rotation of the particle. Moreover, dynamical stability of the system can be achieved only when minor rotational perturbations lead to the emergence of a restorative torque in the opposite direction, as depicted in Fig.~\ref{fig:Rayleigh_acoustomechanics}(a). For the orientation angle $\theta$ under consideration and the corresponding $z$-directed torque, this means $T_z = 0$ and $\partial T_z / \partial \theta < 0$.

\subsection{The Rayleigh approximation}
\label{sec:Rayleigh_torque}

In the Rayleigh approximation $\{ka,kb\}\ll 1$, for a spheroid particle illuminated by an $x$-propagating plane wave, Eq.~\eqref{eq:torque_corr} yields $\vb{T} = T_z \bar{\vb{z}}$,
\begin{equation}
\label{rayleigh-torque}
T_z = -\frac{\beta \abs{p_0}^2}{4} \sin{2\theta} \left(\alpha^{\text{st}}_{\text{D},s} - \alpha^{\text{st}}_{\text{D},l} \right) \left( 1 + \frac{k^6}{72 \pi^2} \alpha^{\text{st}}_{\text{D},s} \alpha^{\text{st}}_{\text{D},l} \right).
\end{equation}
The equation extends the approximate torque expression obtained in Ref.~\cite{Fan2008JASA} from heuristic arguments. 
The torque \eqref{rayleigh-torque} has the same $\propto \sin 2\theta$ dependence as the lateral force $F_y$, and a nontrivial tilted-particle configuration with vanishing torque can be achieved only for $\alpha^{\text{st}}_{\text{D},s} \alpha^{\text{st}}_{\text{D},l} = - 72 \pi^2 / k^6$. 
However, it follows from Eq.~(\ref{eq:static_monopole_dipole}) and the fact that $L_i < 1$ that for a particle with isotropic effective mass density the elements of polarizability tensor always have the same sign and this condition cannot be satisfied, and hence there is no stable acoustic lift for a spheroid particle in the Rayleigh regime.

We performed numerical calculations of the acoustic force and torque on a spheroid silicone rubber particle with parameters $kb = 0.01, ka = 0.002$ using Comsol Multiphysics. The results are shown in Figs.~\ref{fig:Rayleigh_acoustomechanics}(b,c). One can see perfect agreement with the analytical expressions in the Rayleigh regime. The only stable point with $T_z=0$ and $\partial T_z / \partial \theta < 0$ is $\theta = 0$, when the spheroid is oriented perpendicular to the incident field, and the lateral force vanishes. The acoustic radiation torque in this and all subsequent plots is normalized by $T_0 = F_0/k = \sigma_{\text{geom}} \beta \abs{p_0}^2 / 2 k$.

\subsection{Beyond the Rayleigh approximation}
\label{subsec:resonant_particle}

Since the stable optical lift was achieved for structures much larger than the wavelength \cite{optical-lift-nature, optical-lift-optica}, one can expect that acoustic lift effect may also be discovered beyond the Rayleigh limit. In this section, we extend our analysis to the Mie-resonant scattering. 
For numerical calculations, we choose a highly resonant prolate spheroid with $a/b = 0.2$, material parameters $\bar{\beta} = 3$, $\bar{\rho} = 7$, and variable parameter $ka$.
The Rayleigh-limit expression for the torque, Eq.~\eqref{rayleigh-torque}, is not valid for larger particles because of: 
(i) higher multipole contributions and
(ii) nonlocal effects for each spherical multipole since spheroid has lower symmetry compared to sphere \cite{Tsimokha2022Apr}.

To address this problem we employ the well-established partial-wave expansion~\cite{blackstock,williams}. First, we decompose the incident and scattered fields into spherical multipoles:
\begin{equation}
\label{multipolar-decomposition}
p_{^{\text{inc}} _{\text{sc}}} (r, \varphi, \vartheta) = p_0 \sum \limits_{nm} \:^{A_{nm}} _{S_{nm}} R_n (kr) Y_n^m (\varphi, \vartheta),
\end{equation}
where $\sum_{nm} = \sum_{n = 0}^{\infty} \sum_{m = -n}^{n}$, $A_{nm}$ and $S_{nm}$ are the expansion coefficients of the incident and scattered fields, $R_n$ are the radial  functions which are equal to spherical Hankel (Bessel) functions for the scattered (incident) field, and $Y_n^m$ are the spherical harmonics of order $n$ and with the azimuthal number $m$ correspondent to the angular momentum about the $z$-axis. 
The torque can be expressed from the exact equation \eqref{force_tensor} taking all excited multipoles into account~\cite{Gong2021May}:
\begin{equation}
T_z = \sum \limits_{nm} T_{z}^{(nm)} =  \frac{\beta \abs{p_0}^2}{2 k^3} \Re \sum \limits_{nm} m \left(A_{nm}^{*} + S_{nm}^{*} \right) S_{nm}.
\label{eq:tz_all_multipoles}
\end{equation}
This equation is valid for arbitrary particle and arbitrary incident field. All information about the particle and the incident field is encoded in the the coefficients $S_{nm}$ and $A_{nm}$, respectively.

The total dipole contribution to the torque (including nonlocal effects) is given by the $n=1$ terms in Eq.~\eqref{eq:tz_all_multipoles}: $T_{z,\text{D}} = \sum \limits_{m} T_z^{(1m)}$.
This dipole torque can still be expressed via Eq.~\eqref{eq:torque_corr} using the total induced dipole moment $\vb{D}$. 
However, due to the nonlocal effects, the excited dipole moment can no longer be expressed through the local velocity ${\bf v}$, i.e.,
Eq.~\eqref{monopole_dipole_definition} is no longer valid. Instead, the general nonlocal relation between the induced dipole moment and the incident acoustic field takes on the form~\cite{Bobylev2020Sep}:
\begin{equation} 
\label{eq:dipole_decomposition}
\vb{D} = 
{\boldsymbol{\chi}}^{\text{D}} p + \ten{\alpha}^{\text{D}} \vb{v} + {\ten{\beta}^{\text{D}}} \grad \vb{v} + \ten{\gamma}^{\text{D}} \grad \grad \vb{v} + \dots \,,
\end{equation}
where \{${\boldsymbol{\chi}}^{\text{D}}$, ${\ten{\alpha}^{\text{D}}}$, ${\ten{\beta}^{\text{D}}}$, ${\ten{\gamma}^{\text{D}}}$\} are tensors of the first, second, third and fourth rank, respectively, which depend on the particle's parameters and orientation.
These tensors describe the coupling between the \{monopole, dipole, quadrupole, octupole\} modes in the incident field and the dipole mode of the particle. For instance, the first term in Eq.~\eqref{eq:dipole_decomposition} is known as the Willis coupling \cite{willis-1,willis-2,alu_willis,Sepehrirahnama2022Oct,Quan2018Jun,Sieck2017Sep,Willis1985Jan}.
Depending on the particle's symmetry some of the coefficients in the decomposition~\eqref{eq:dipole_decomposition} may vanish. 
In particular, from the group-theory analysis of the scattered multipolar contributions one can conclude that ${\boldsymbol{\chi}}^{\text{D}} \equiv {\bf 0}$ and ${\ten{\beta}^{\text{D}}} \equiv {\bf 0}$ for a spheroid and any particle possessing inversion symmetry \cite{Poleva2023Jan, Tsimokha2022Apr}. 
More details on the decomposition \eqref{eq:dipole_decomposition} and individual contributions can be found in the Appendix~\ref{app:nonlocality}.

\begin{figure}[t]
\includegraphics[width=0.86\linewidth]{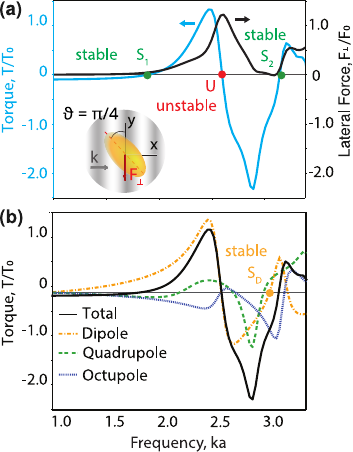}
\caption{\label{fig:Mie_acoustomechanics} Acoustic torque and lateral force on a prolate spheroid particle with eccentricity $e = 0.98$ oriented at the angle $\theta = \pi/4$ vs. the dimensionless particle size $ka$ in the Mie regime. The relative material parameters are $\bar{\beta} = 3$ and $\bar{\rho} = 7$. (a) The acoustic torque $T_z$ exhibits two stable zeros $S_{1,2}$ and unstable zero $U$. The lateral force $F_{\perp}$ is nonzero in these points, enabling acoustic lift (see also Fig.~\ref{fig:color_maps}). (b) Multipolar contributions to the net acoustic torque, which shows stable lift point $S_{\text{D}}$ (corresponding to $S_2$) already in the dipole approximation including higher-order nonlocal contributions to the dipole moment, Eq.~\eqref{eq:dipole_decomposition}.}
\end{figure}

\begin{figure}[t!]
\centering
\includegraphics[width = 1.0\linewidth]{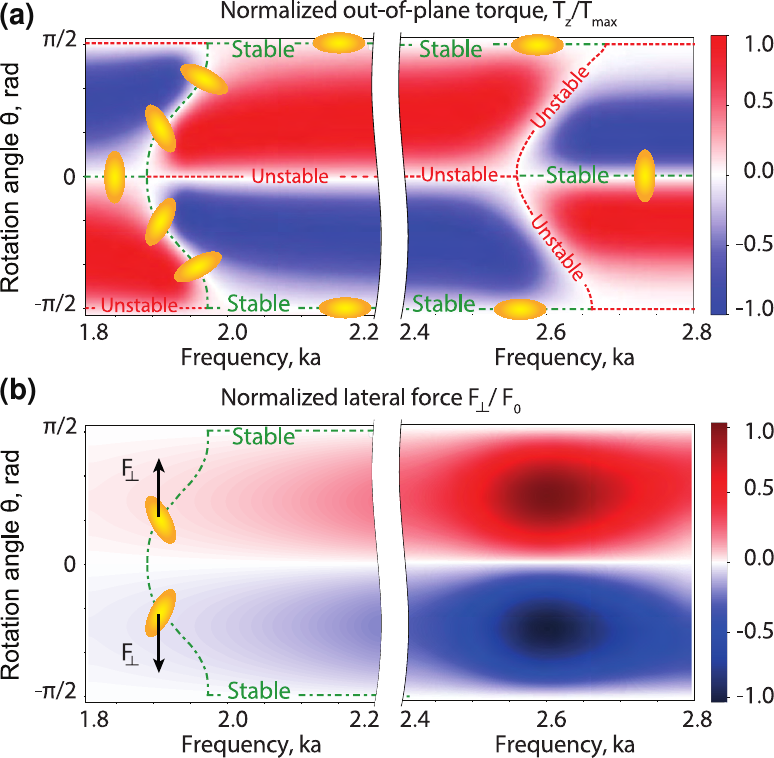}
\caption{\label{fig:color_maps} Color maps of the acoustic torque $T_z$ (a) and lateral force $F_\perp$ (b) on a prolate spheroid particle in the Mie regime vs. the dimensionless particle size $ka$ and orientation $\theta$. The torque is normalized by $T_{\text{max}}$, which is the absolute value of the maximum torque for each value of $ka$. The rotationally stable configurations of the particle correspond to the lines with ${T}_z = 0$ and $\partial T_z / \partial \theta  < 0$. The rotationally stable configuration in (a) with nonzero lateral force in (b) provide a stable acoustic lift.}
\end{figure}

Thus, the total torque in the Mie regime has contributions from all multipole amplitudes. Figure~\ref{fig:Mie_acoustomechanics}(a) shows the numerically calculated acoustic force and torque on a Mie-resonant spheroid particle oriented at $\theta = \pi/4$ versus the $ka$ parameter. In sharp contrast to the Rayleigh regime, the torque vanishes, $T_z=0$, for some values of $ka$, while the lateral force is non-zero. Further analysis (see Appendix~\ref{app:angular_dependence_torque}) shows that the points $S_1$ ($ka = 1.91$) and $S_2$ ($ka=3.12$) correspond to the stable lift configuration with $\partial T_z / \partial \theta < 0$.

The multipolar origin of these stable zero-torque points can be recognized by analysing the individual multipole contributions ($n=1,2,3,...$), shown in Fig.~\ref{fig:Mie_acoustomechanics}(b). One can see that the first stability point $S_1$ appears away from multipolar resonances due to the mutual cancellation of different multipole contributions. In turn, the second stability point $S_2$ is close to the multipolar resonance, where not only the sum of mutlipoles contribution vanishes, but also each multipole contribution vanishes in the vicinity of this point. Such stability point appears already in the dipole response ($n=1$) taking into account nonlocal effects, Eq.~\eqref{eq:dipole_decomposition} (see Appendix~\ref{app:nonlocality}).

To describe the dependence of the stable acoustic lift on the orientation angle $\theta$, in Fig.~\ref{fig:color_maps} we plot the acoustic torque and lateral force as dependent on $ka$ (in the range including the stable zero-torque point $S_1$ and unstable point $U$) and $\theta$. One can see that these zero-torque points evolve smoothly with $\theta$, so that the stable acoustic lift can appear at any orientation of the particle.

\section{Example: Sorting of blood cells}
\label{sec:discussion}

We briefly discuss a potential application of the acoustic lateral force and lift for acoustic sorting of particles with different shapes~\cite{Khan2024MicrosystNanoeng}. As a proof-of-principle demonstration, we choose the common blood cell separation scenario \cite{Urbansky2017Dec}, namely the separation of near-spherical leukocytes (white blood cells) and axisymmetric erythrocytes (red blood cells). This section serves only as a conceptual illustration of the acoustic shape-based sorting; the current prevalent methods of acoustic or microfluidic sorting of particles are based on their size and material parameters \cite{Fan2022Sep,Rufo2022Apr,Wu2019Jun,Olofsson2020, Ali2024Sep}.  

\begin{figure}[h!]
\includegraphics[width=1.0\linewidth]{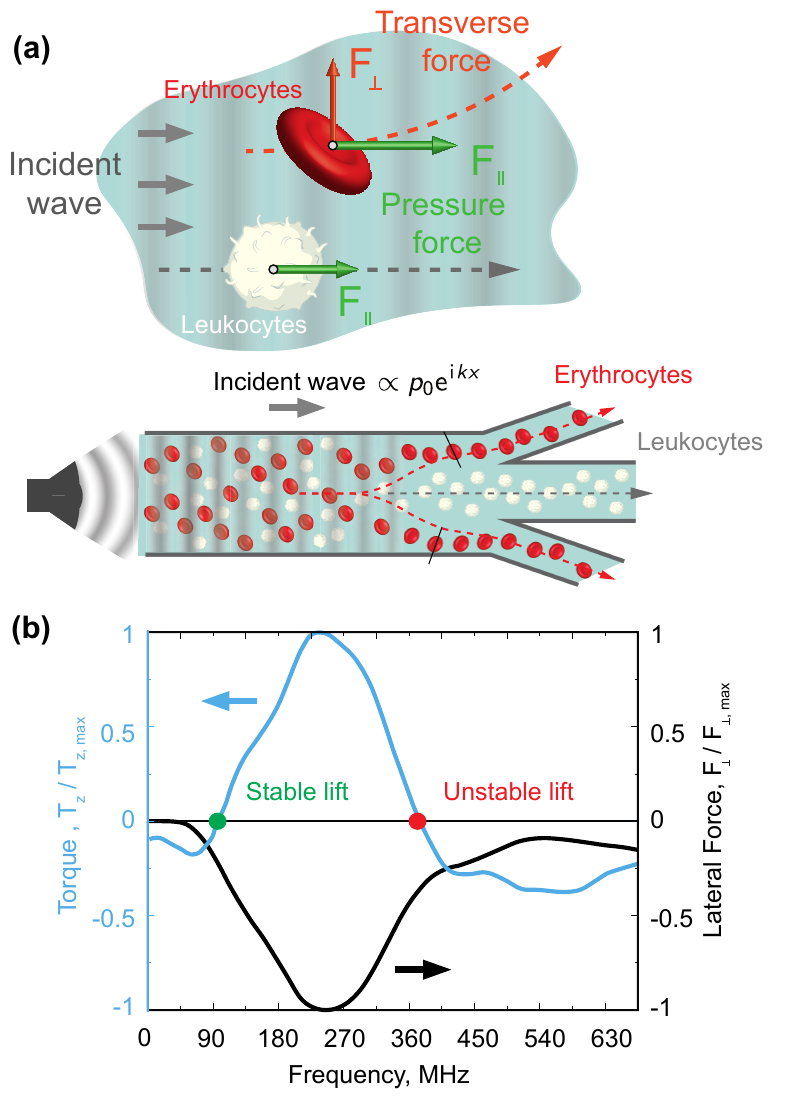}
\caption{\label{fig:lift_separation}
(a) Schematics of the possible application of acoustic lift for acoustofluidic separation of the white and red blood cells using their shape anisotropy. (b) Frequency dependences of the acoustic lateral force and torque exerted by a plane-wave on a red blood cell particle in blood plasma at the orientation angle $\theta = \pi/4$. The values are normalized by the corresponding maximum absolute values. There is a point of stable lift at $ka \simeq 2$ which corresponds to the frequency of about $90$~MHz.}
\end{figure}

The main idea is depicted in Fig.~\ref{fig:lift_separation}(a). An acoustic wave propagates along the microchannel containing red and white blood cells. It produces lateral forces (with different signs for different orientations) on the red blood cells due to their elongated shapes, while spherical white blood cells experience only the longitudinal pressure force. The three-channel geometry, often used in microfluidic systems, may be used to implement the sorting, as shown in Fig.~\ref{fig:lift_separation}(b). 

We calculate acoustic force and torque exerted by a plane wave incident on a red blood cell submerged in blood plasma. The red blood cell geometry is modelled as a biconcave solid disc described by the implicit equation from~\cite{Evans1972Oct}. The size and shape parameters are taken from the table in \cite{silva}. The acoustic and mechanical parameters of erythrocytes and blood plasma are taken from~\cite{Gupta2018Apr, Dulinska2006Mar}. For simplicity, we neglect thermoviscous effects.   

A red blood cell has the same symmetry as a spheroid, so that our previous analysis can be applied. The results of numerical calculations are shown in Fig.~\ref{fig:lift_separation}(c). The frequency dependence of the acoustic torque exhibits a stable-lift point with $\omega /2\pi \simeq 90$~MHz corresponding to $ka \simeq 2$. For this frequency, the ratio of the lateral-to-longitudinal force components is $|F_\perp / F_\parallel | \simeq 0.2$. Assuming the incident pressure wave amplitude $p_0 = 10$~kPa, the lateral force is $|F_\perp | \simeq 8$~fN, which is significant for the red blood cell weight $m\simeq 30$~picograms \cite{Phillips2012Sep}.

\section*{Conclusion}
\label{sec:conclusion}

We have examined acoustic forces and torque on an anisotropic (spheroid) particle in an incident plane-wave field. Within the monopole-dipole approximation, we have derived exact analytical expressions, including previously unknown recoil contributions.  
Similar recoil terms are well studied in optical forces and torques~\cite{chaumet,nieto-optics-express,optical-pulling-ct-chan,Bliokh2014Mar,Bekshaev2015Mar,Antognozzi2016Aug,Yevick2016Apr,Hayat2015Oct}. Importantly, the acoustic recoil force on an anisotropic particle can have lateral component, orthogonal to the incident-wave direction. The sign of this lateral force is associated with the relative phases between the induced monopole and dipole and can be controlled by physical and geometric parameters of the particle.

We have also examined acoustic torque on a spheroid particle in the presence of the lateral force. In particular, we have found conditions for the rotationally stable lateral drift of the particle, i.e., stable acoustic lift. The optical counterpart of this effect has been described in~\cite{optical-lift-nature, optical-lift-optica}. Notably, the stable acoustic lift does not appear in the Rayleigh limit and requires larger Mie particles involving higher-order terms in the scattering problem.

Finally, we propose a concept for acoustofluidic sorting of particles according to their shapes, using an example of the red and white blood cell separation. 
We have shown that there exists a set of parameters where such sorting could potentially be implemented using the considered acoustic lift effect.
This is a proof-of-principle consideration, as creating such a travelling-wave acoustic setup can be unnecessarily challenging, and a more detailed consideration including thermoviscous and acoustic streaming effects    \cite{Muller2013Aug, Nama2015Jun, Hsu2020Sep} is required for practical implementation. Furthermore, for dense solutions multiple re-scattering should be taken into account~\cite{Zhang2016}. 
Nonetheless, the proposed approach can be further developed as a useful tool.

Thus, our results reveal several new acoustomechanical phenomena, which can provide novel functionalities for acoustic and microfluidic manipulation of particles. To further explore these phenomena, it would be important to examine recoil forces and torques in structured incident fields~\cite{Bliokh2023Aug}, such as evanescent waves~\cite{PRL, Wei2020Aug,Long2020Jun} or vortex beams~\cite{Zhang2011Dec, Zhang2020Jun, Muelas-Hurtado2022Nov,Guo2022Dec}, as well as for structured particles made of nontrivial inhomogeneous or elastic materials~\cite{kotelnikova_determination_2021}.

\section*{Materials and Methods}

\subsection*{Numerical simulation of acoustic scattering and Acoustic Radiation Force and Torque}
The acoustic wave and acoustic force simulation was performed using a finite element method in Linearized Acoustic module of COMSOL Multiphysics~5.5. To simulate the properties of free-space propagation, PML-layer surrounding a particle was used. The particle parameters used are outlined in the main text. The radiation patterns were calculated using built-in COMSOl Radiation pattern feature.

The forces and torques were calculated by integration of the explicit definition of acoustic stress tensor in Eq.~\eqref{force_tensor} over a sphere enclosed around a particle.

Due to high sensitivity to mesh size, the Acoustic Radiation Force and Torque in Rayleigh regime were calculated using a 2D-axisymmetric model, the details of calculation are outlined in Appendix~\ref{app:numerical_axisymmetric}.

The multipole scattering coefficients $S_{nm}$ in the decomposition (Eq. \eqref{multipolar-decomposition})
can be retrieved using integration over sphere in the following way:
\begin{equation}
    S_{nm} =  \frac{1}{h_n^{(2)} (k R_{\text{int}})} \frac{1}{R_{\text{int}}^2} \int \limits_{\partial S} \dd S  Y_n^{m*} (\theta, \phi) p_s (R_{\text{int}}, \theta, \phi) / p_0 . 
\end{equation}
which is possible due to orthogonality of spherical harmonics on a spherical surface. Note that Hankel function of the second kind $h_n^{(2)}$ is used because COMSOL Multiphysics employs the $e^{+i \omega t}$ convention for time-harmonic fields. 

The scattering cross section was also calculated in 3D first by retrieving scattered energy, and dividing it by incident acoustic intensity:
\begin{equation}
    \sigma_{\mathrm{sc}} = \frac{W_{\text{sc}}}{I_{\text{inc}}} = \frac{1}{I_{\text{inc}}} \int \limits_{\partial S} \dd S \, \frac{1}{2} \Re \left[ (\mathbf{n} \cdot \mathbf{v_s}) p_s^* \right],
\end{equation}
where $I_{\text{inc}} = \frac{1}{2} \Re \left[ p_i^* v_i  \right] = \frac{1}{2} \abs{p_0}^2  / (c_{\text{s}} \rho)$. 

\begin{acknowledgments}
The authors are grateful to Kristina Frizyuk for valuable discussions. This work was supported by Russian Academic Leadership Programm Priority 2030. The numerical simulations were supported by Russian Science Foundation (grant No. 20-72-10141). 
K.Y.B. acknowledges support from 
Ikerbasque (Basque Foundation of Science), 
Marie Sk\l{}odowska-Curie COFUND Programme of the European Commission (project HORIZON-MSCA-2022-COFUND-101126600-SmartBRAIN3), 
the International Research Agendas Programme (IRAP) of the Foundation for Polish Science co-financed by the European Union under the European Regional Development Fund and Teaming Horizon 2020 program of the European Commission [ENSEMBLE3 Project No. MAB/2020/14], 
and the project of the Ministry of Science and Higher Education (Poland) ``Support for the activities of Centers of Excellence established in Poland under the Horizon 2020 program'' [contract MEiN/2023/DIR/3797],

\end{acknowledgments}

\bibliography{refs.bib}

\onecolumngrid
\newpage
\appendix

\section{Derivation of the recoil force}
\label{app:force_derivation}

Here we derive the recoil corrections to the main monopole 
 and dipole forces, Eq.~(\ref{eq:full_force}). The recoil correction to the torque, Eq.~(\ref{eq:torque_corr}), can be derived in a similar way. 

We start with the general expression for the acoustic force:
\begin{equation}
    \label{tensor_decomposition}
    \mathbf{F} = - \oint \left( \ten{\mathcal{T}}_{\text{inc}} + \ten{\mathcal{T}}_{\text{mix}} + \ten{\mathcal{T}}_{\mathrm{recoil}} \right)\cdot \dd \vb{S}\,,
\end{equation}
where $\ten{\mathcal{T}} = \frac{1}{4} \ten{I} \left( \beta \abs{p}^2 - \rho \abs{\vb{v}}^2 \right) + \frac{1}{2} \rho \Re\left( \vb{v}^* \otimes \vb{v} \right)$, where $\vb{v}^*\otimes \vb{v}$ is the outer product of two vectors, and the total field is the sum of the incident and scattered field: $(p, \vb{v}) = (p_{\text{inc}}, \vb{v}_{\text{inc}}) + (p_{\text{sc}}, \vb{v}_{\text{sc}})$.

The pressure fields produced by the monopole and dipole moments located at $\mathbf{r}_0$ can be written as~\cite{williams}:
\begin{equation}
\label{scattered_pressure_monopole}
p_{\mathrm{M}} = -\rho \omega^2 M \frac{\eu^{\iu k r}}{4 \pi r},
\end{equation}
\begin{equation}
\label{scattered_pressure_dipole}
p_{\mathrm{D}} = -\iu \rho c k (\mathbf{D} \cdot \mathbf{n}) \frac{\iu k r - 1}{4 \pi r^2} \eu^{\iu k r}.
\end{equation}
where $M$ and $\vb{D}$ are the monopole and dipole strengths, respectively. 
The corresponding monopole and dipole velocity fields are obtained from the Euler equation ${\bf v} = -i/(\rho \omega) {\nabla} p$:
\begin{equation}
\label{scattered_velocity_monopole}
\mathbf{v}_{\mathrm{M}} = \frac{-\iu \omega M}{4 \pi} 
\vb{n}  \frac{1 - \iu kr}{r^2} \eu^{\iu k r}, 
\end{equation}
\begin{equation}
    \label{scattered_velocity_dipole}
    \mathbf{v}_{\mathrm{D}} = 
    - \frac{\eu^{\iu k r}}{4 \pi r^3} \left[\mathbf D \left(\iu k r - 1\right) - \mathbf{n} (\mathbf{D} \cdot \mathbf{n}) \left( k^2r^2 + 3\iu kr -3\right) \right],
\end{equation}
where $\vb{n} = \vb{r}/r$.

The mixed terms $\ten{\mathcal{T}}_{\mathrm{mix}}$ in Eq. (\ref{tensor_decomposition}) can be integrated in the {\it near} field using the gradient expansion of the incident field near the position of the particle. Retaining the monopole and dipole near-field contributions, Eqs.\eqref{scattered_pressure_monopole}, \eqref{scattered_pressure_dipole}, \eqref{scattered_velocity_monopole}, \eqref{scattered_velocity_dipole}, we obtain the previously-known expression for the pure monopole and dipole force contributions \cite{bruus-7, powell, PRL}:
\begin{equation}
\label{no_recoil}
    \mathbf{F} =  \frac{1}{2} \Re \left[  M^* \nabla p \right] + \frac{\rho}{2} \Re \left[ \vb{D}^* \cdot (\nabla) \mathbf{v}  \right] = \mathbf{F}_{\mathrm{M}} + \mathbf{F}_{\mathrm{D}}\,.
\end{equation}
Next, the \textit{self-action} of the scattered field, described by the term $\ten{\mathcal{T}}_{\mathrm{recoil}}$ in Eq.~(\ref{tensor_decomposition}), is usually neglected for small Rayleigh particles because the corresponding force scales as $\propto (ka)^6$. 
Here we derive this recoil force contribution:
\begin{equation}
\label{self_force_tensor}
\mathbf{F}^{\mathrm{recoil}}  = - \oint   \ten{\mathcal{T}}_{\text{recoil}}  \cdot \dd \mathbf{S} = - \oint \left[ \left(\frac{\beta}{4} | p_{\text{sc}} |^2 - \frac{\rho}{4} | \mathbf{v}_{\text{sc}}|^2 \right) \hat{\mathbf{I}} + \frac{1}{2}\rho  \mathbf{v}_{\text{sc}}^{*} \otimes \mathbf{v}_{\text{sc}}  \right] \cdot \dd \mathbf{S}.
\end{equation}
It is  easier to integrate this expression in the {\it far} field, $kr \to \infty$, where it is reduced to \cite{Westervelt1951}:
\begin{equation}
\label{self_force_tensor_far_field}
\mathbf{F}^{\mathrm{recoil}} =   - \frac{\rho}{2} \Re \oint \mathbf{v}^*_{\text{sc}} (\mathbf{v}_{\text{sc}} \cdot \dd \mathbf{S}).
\end{equation}
The monopole and dipole velocity far fields \eqref{scattered_velocity_monopole} and \eqref{scattered_velocity_dipole}, are:
\begin{align}
\label{v_far_field}
\mathbf{v}_{\mathrm{M}} \simeq - \frac{ \omega M}{4 \pi} \frac{k}{r} e^{\iu k r} \mathbf{n}\,, \quad
\mathbf{v}_{\mathrm{D}}  \simeq \frac{e^{i k r}}{4 \pi r} k^2 \mathbf{n} (\mathbf{D} \cdot \mathbf{n})\,.
\end{align}
Substituting the scattered velocity field as the sum of the monopole and dipole parts,  
$\mathbf{v}_{\text{sc}} = \mathbf{v}_{\mathrm{M}} + \mathbf{v}_{\mathrm{D}}$, into Eq.\eqref{self_force_tensor_far_field}, we obtain for the force components:
\begin{align}
\label{correction_through_multipole_velocities}
F_i^{\mathrm{recoil}} =  
- \frac{\rho}{2} \, \mathrm{Re} \oint \dd S \left[ v^{*i}_{\mathrm{M}} (v^{j}_{\mathrm{M}} n_j) + v^{*i}_{\mathrm{D}} (v^{j}_{\mathrm{M}} n_j) + v^{*i}_{\mathrm{M}} (v^{j}_{\mathrm{D}} n_j) + v^{*i}_{\mathrm{D}} (v^{j}_{\mathrm{D}} n_j)\right],
\end{align}
where we used $d{\bf S} = {\bf n} dS $.
Using the identities for the solid-angle integration $ \oint \dd S = r^2 \int\limits_{4 \pi} \dd \Omega$,
\begin{align}
    \oint\limits_{4 \pi} \dd \Omega \, n_i = 0, \quad \int\limits_{4 \pi} \dd \Omega \, n_i n_j = \frac{4 \pi}{3} \delta_{ij}, \quad \int\limits_{4 \pi} \dd \Omega \, n_i n_j n_k = 0 
\end{align}
the first and the last terms of Eq.~\eqref{correction_through_multipole_velocities} vanish:
\begin{align}
    \int\limits_{\partial S_\infty} \dd S \; v^{*i}_{\mathrm{M}} (v^{j}_{\mathrm{M}} n_j) & \propto \int\limits_{4 \pi} \dd \Omega \, n_i n_j n_j = 0\,, \\
    \int\limits_{\partial S_\infty} \dd S \; v^{*i}_{\mathrm{D}} (v^{j}_{\mathrm{D}} n_j) & \propto \int\limits_{4 \pi} \dd \Omega \, n_i n_l n_k n_j n_j = \int\limits_{4 \pi} \dd \Omega \, n_i n_l n_k  = 0\,.
\end{align}
In turn, the middle terms of Eq.~\eqref{correction_through_multipole_velocities} yield:
\begin{align}
    r^2 \int\limits_{4 \pi} \dd \Omega \,  v^{*i}_{\mathrm{D}} (v^{j}_{\mathrm{M}} n_j) &= - \frac{1}{4 \pi} \, k^2 \, \frac{\omega M}{4 \pi} k \int\limits_{4 \pi} \dd \Omega \, n_i D^*_k n_k  = -\frac{k^3}{12 \pi} \omega M D^*_i, \\
    r^2 \int\limits_{4 \pi} \dd \Omega \,  v^{*i}_{\mathrm{M}} (v^{j}_{\mathrm{D}} n_j) &= - \frac{1}{4 \pi} \, k^2 \,  \frac{\omega M^*}{4 \pi} k \int\limits_{4 \pi} \dd \Omega \, n_i D_k n_k =  - \frac{k^3}{12 \pi} \omega  M^* \,  D_i.
\end{align}
This results in the final expression for the recoil force:
\begin{align}
\label{self_force}
    \vb{F}^{\mathrm{recoil}} = - \frac{ k^4}{12 \pi} \sqrt{\frac{\rho}{\beta}} \Re \left( M^* \vb{D} \right).
\end{align}

\section{Magnitude of Recoil force}
\label{app:magnitude}

For a Rayliegh particle of a characteristic size $a$, $ka \ll 1$, the dependence of all terms in the multipole decomposition of the force can be distinguished by the power law dependence~\cite{Bekshaev2013JO,Bekshaev2011AOT}:
\begin{equation}
\log |\vb{F}_i| = \nu_i \log(ka) + \operatorname{const}\,,
\end{equation}
where integer $\nu_i$ specifies the power dependence order, and $i$ is the term label (e.g., ``M'' for the monopole, ``D'' for the dipole, etc.). Number $\nu_i$ depends on the particle and wave properties. The order of the power dependence for a small spherical particle can be found via the Taylor expansion of the corresponding Mie coefficients, see SM of \cite{PRL}.

\begin{table}[!h]
\caption{The leading power-law dependence of the acoustic radiation force for different multipole orders characterized by the particle and wave type.}
\label{table:magnitude}
\begin{tabular}{p{0.05\textwidth} p{0.35\textwidth} p{0.14\textwidth} p{0.14\textwidth} p{0.14\textwidth} p{0.14\textwidth}}
 \hline
No & Scenario & $\mathbf{F}_{\rm{M}}$ & $\mathbf{F}_{\rm{D}}$ &$\mathbf{F}_{\rm{recoil}}$ & $\mathbf{F}_{\rm{Q}}$ \\ [0.5ex] 
 \hline\hline
 1 & Lossless particle, plane wave & $\propto (ka)^6$ & $\propto (ka)^6$ & $\propto (ka)^6$ & $\propto (ka)^{10}$ \\ 
 \hline
 2 & Lossless particle, inhomogeneous wave & $\propto (ka)^3$ & $\propto (ka)^3$ & $\propto (ka)^6$ & $\propto (ka)^5$ \\
 \hline
 3 & Lossy particle, plane wave & $\propto (ka)^3$ & $\propto (ka)^3$ & $\propto (ka)^6$ & $\propto (ka)^5$ \\
 \hline
4 & Lossy particle, inhomogeneous wave & $\propto (ka)^3$ & $\propto (ka)^3$ & $\propto (ka)^6$ & $\propto (ka)^5$ \\
 \hline
 \hline
\end{tabular}
\end{table}

It can be seen from Table~\ref{table:magnitude} that for a lossless particle in a plane-wave field the recoil force is of the same order as the main monopole and dipole components. Figure~\ref{fig:parameter_map} shows that the lateral recoil force can reach substantial percentage of the total force acting on the particle. 
Note that all forces in Table~\ref{table:magnitude} are not normalized and actually scale as $\propto (ka)^\nu / k^2$. In the main text results are normalized by $F_0 = \sigma_{\text{geom}}\beta |p_0|^2 / 2$ with $\sigma_{\text{geom}} \propto a^2$ being the geometrical cross section.  Therefore, the main-text results depend on the dimensionless parameter $ka$, and scale as $F/F_0 \propto (ka)^{\nu-2}$.

\section{Polarizability derivation}
\label{app:polarizability}

\subsection{Rayleigh limit}

Here we derive the monopole and dipole polarizabilities of a small  ellipsoid particle. The semi-axes along the $x$, $y$, and $z$ axes are assumed to equal $a$, $b$, and $c$, respectively.
The \textit{monopole} moment is proportional to a volume flow through the object per second \cite{blackstock}, and in the lowest-order approximation the monopole polarizability should only depend on the relative compressibility $\bar{\beta} = {\beta_{\text{p}}}/{\beta}$ and the particle's volume. Taking the polarizability of a sphere, one can renormalize it for the ellipsoid volume:
\begin{equation}
    \alpha_{\text{M}}^{\text{sphere}} = \frac{4 \pi}{3} a^3 (\bar{\beta} - 1) \quad \rightarrow \quad 
    \alpha_{\text{M}}^{\text{ellipsoid}} = \alpha_{\text{M}}^{\text{st}} = \frac{4 \pi}{3} a b c  (\bar{\beta} - 1).
    \label{static_monopole}
\end{equation}

To derive the \textit{dipole} polarizability, it is instructive to consider the analogy with optics, since the electric-dipole polarizability for anisotropic particles is well known \cite{correction-2, electric-anisotropy2}. Our derivation of the acoustic dipole polarizability closely follows optical textbook \cite[\S~5.3]{bohren-huffman}.
Acoustic pressure field obeys the Helmholtz equation. For small Rayleigh particles, we can use  the `quasistatic' limit $c_s^{-2} \partial_t^2 p \ll \nabla^2 p$ resulting in the Laplace equation:
\begin{equation}
    \nabla^2 p - \frac{1}{c_{\mathrm{s}}^2} \frac{\partial^2 p}{\partial t^2}=0 \quad \longrightarrow \quad  \nabla^2 p \simeq 0\,.
    \label{eq:Laplace}
\end{equation}
Using the ellipsoidal coordinates $(\lambda, \mu, \nu)$ with the value ranges $-c^2 < \lambda < \infty$, $-b^2 < \mu < - c^2$, $-a^2 < \nu < -b^2$, the Laplace equation \eqref{eq:Laplace} becomes:
\begin{equation}
    (\mu - \nu) f(\lambda) \frac{\partial}{\partial \lambda} \left[ f(\lambda) \frac{\partial p}{\partial \lambda} \right] + (\nu - \lambda) f (\mu) \frac{\partial}{\partial \mu} \left[ f(\mu) \frac{\partial p}{\partial \mu} \right] +  (\lambda - \mu) f(\nu) \frac{\partial}{\partial \nu} \left[ f(\nu) \frac{\partial p}{\partial \nu} \right]=0,
    \label{laplace_ellipsoidal}
\end{equation}
where $f(q) = \sqrt{(q+a^2) (q+b^2) (q+c^2)}$. The connection between $(x,y,z)$ and $(\lambda, \mu, \nu)$ is
\begin{equation}
    x^2 = \frac{(a^2 + \lambda)(a^2 + \mu) (a^2 + \nu)}{(b^2 - a^2)(c^2 - a^2)}, \quad 
    y^2 = \frac{(b^2 + \lambda)(b^2 + \mu) (b^2 + \nu)}{(b^2 - a^2)(c^2 - a^2)}, \quad 
    z^2 = \frac{(c^2 + \lambda)(c^2 + \mu) (c^2 + \nu)}{(b^2 - a^2)(c^2 - a^2)}.
    \label{eq:lambda_mu_nu}
\end{equation}
Next, we consider an ellipsoid in a quasistatic velocity field $\vb{v}_{\text{inc}} = v_0 \bar{\vb{z}}$, which corresponds to the pressure field
\begin{equation}
    p_{\text{inc}} = i \omega \rho v_0 z = i \omega \rho v_0 \left[\frac{(c^2 + \lambda)(c^2 + \mu) (c^2 + \nu)}{(b^2 - a^2)(c^2 - a^2)} \right]^{1/2}.
\end{equation}
On the boundary of the particle ($\lambda = 0$) the pressure and normal component of the velocity are required to be continuous, while at the infinity ($\lambda \to \infty$) the scattered field must vanish:
\begin{subequations}
\begin{align}
    \label{eq:bc_p}
    p_{\text{inside}}(0, \mu, \nu) &= p_{\text{inc}}(0, \mu, \nu) + p_{\text{sc}}(0, \mu, \nu), \\
    \label{eq:bc_v}
    v_{\text{inside},\lambda}(0, \mu, \nu) &= v_{\text{inc},\lambda}(0, \mu, \nu) + v_{\text{sc},\lambda}(0, \mu, \nu) \quad \to \quad \frac{\rho}{\rho_{\text{p}}} \frac{\partial}{\partial \lambda} p_{\text{inside}} = \frac{\partial}{\partial \lambda} p_{\text{inc}} + \frac{\partial}{\partial \lambda} p_{\text{sc}}, \\
 p_{\text{sc}}(\infty, \mu, \nu) &= 0,
    \label{eq:bc_infty}
\end{align}
\end{subequations}
where $p_{\rm inside}$ and ${\bf v}_{\rm inside}$ are the fields inside the particle.
Based on the symmetry of the problem, we search the solution for $p_{\text{sc}}$ and $p_{\text{inside}}$ in the form \cite{bohren-huffman}
\begin{equation}
p(\lambda, \mu, \nu) = F(\lambda) \left[ (c^2 + \mu) (c^2 + \nu^2) \right]^{1/2}.
\end{equation}
Substituting this to the Laplace equation \eqref{laplace_ellipsoidal}, one can find that $F(\lambda)$ satisfies 
\begin{equation}
    f(\lambda) \frac{\dd}{\dd \lambda} \left[ f(\lambda) \frac{d F}{\dd \lambda} \right] - \left(\frac{a^2 + b^2}{4} + \frac{\lambda}{2} \right) F(\lambda) = 0. 
\end{equation}
There are two independent solutions~\cite[\S~5.3]{bohren-huffman}:
\begin{equation}
    F_1(\lambda) = \sqrt{c^2 + \lambda}, \qquad 
    F_2(\lambda) = F_1(\lambda) \int \limits_{\lambda}^{\infty} \frac{\dd q}{F_1^2(q)f(q)}.
\end{equation}
Among these two solutions only $F_2(\lambda)$ satisfies Eq.~\eqref{eq:bc_infty}, and hence it should be chosen for $p_{\rm sc}$.
In turn, $F_2(\lambda)$ diverges at the origin, $F_2(-c^2) = \infty$, and $F_1(\lambda)$ should be chosen for the $p_{\text{inside}}$:
\begin{align}
    p_{\text{inside}}(\lambda, \mu, \nu) &= C_{\text{inside}} F_1(\lambda) \left[ (c^2 + \mu) (c^2 + \nu^2) \right]^{1/2}, \nonumber \\
    p_{\text{sc}}(\lambda, \mu, \nu) &= C_{\text{sc}} F_2(\lambda) \left[ (c^2 + \mu) (c^2 + \nu^2) \right]^{1/2}.
    \label{eq:solution_ansatz}
\end{align}
From these equations and Eq.~\eqref{eq:lambda_mu_nu}, one can see that $p_{\text{inside}} \propto  z$, so that the velocity field inside the particle is uniform and parallel to $\vb{v}_{\text{inc}}$. Substituting Eq.~\eqref{eq:solution_ansatz} into the Eqs.~(\ref{eq:bc_p}--\ref{eq:bc_v}), we find 
\begin{equation}
    p_{\text{sc}}(\lambda, \mu, \nu) = p_{\text{inc}}(\lambda, \mu, \nu) \frac{a b c}{2} \frac{\bar{\rho} - 1}{\bar{\rho} + L_3 (1 - \bar{\rho})}\int\limits_\lambda^\infty \frac{\dd q}{(c^2 + q)f(q)}, \qquad 
   L_3 = \frac{a b c}{2} \int\limits_0^\infty \frac{\dd q}{(c^2 + q) f(q)},
    \label{eq:psc_solution}
\end{equation}
where $\bar{\rho} = \rho_{\text{p}}/ \rho$ is the relative mass density, and $L_3$ is the \textit{geometric factor} of the ellipsoid along the $z$ axis.

At large distances $r \gg (a,b,c)$, \eqref{eq:psc_solution} can be approximated as
\begin{equation}
    p_{\text{sc}} \simeq \frac{i \omega \rho v_0 \cos\theta}{r^2} \frac{abc}{3} \frac{\bar{\rho} - 1}{\bar{\rho} + L_3 (1 - \bar{\rho})}.
    \label{eq:psc_solution_far}
\end{equation}
At the same time, since we work in the quasistatic approximation, this distance can still be with the {\it near-field} zone $kr \ll 1$, where the dipole field \eqref{scattered_pressure_dipole} is approximated by 
\begin{equation}
    p_{\text{D}} \simeq \frac{i \rho c k D_z \cos \theta}{4\pi r^2}.
    \label{eq:pD_nf}
\end{equation}
Equaling Eq.~\eqref{eq:psc_solution_far} and Eq.~\eqref{eq:pD_nf}, $p_{\text{sc}} = p_{\text{D}}$, and using the relation $\vb{D} = \ten{\alpha}^{\text{st}}_{\text{D}} \vb{v}$, 
we find the corresponding component of the polarizability tensor:
\begin{equation}
\label{static_dipole_z}
\alpha^{\text{st}}_{d, 3} =\frac{4 \pi}{3}  a b c \frac{\bar{\rho} - 1}{\bar{\rho} + L_3 \left(1 - \bar{\rho} \right)}.
\end{equation}

Considering similar problems for the incident plane waves along the $x$ and $y$ axes, we find the diagonal polarizability tensor of a subwavelength ellipsoid:

\begin{equation}
\label{static_dipole}
    \ten{\alpha}_{\text{D}}^{\text{st}} =  
    \begin{pmatrix}
        \alpha^{\text{st}}_{d, 1} & 0 & 0 \\
        0 & \alpha^{\text{st}}_{d, 2} & 0 \\
        0 & 0 & \alpha^{\text{st}}_{d, 3}
    \end{pmatrix}, \quad 
    \alpha^{\text{st}}_{d, i} =\frac{4 \pi}{3}  a b c \frac{\bar{\rho} - 1}{\bar{\rho} + L_i \left(1 - \bar{\rho} \right)},
\end{equation}
where
\begin{equation}
    L_1 = \frac{a b c}{2} \int\limits_0^\infty \frac{\dd q}{(a^2 + q) f(q)}, \quad  
    L_2 = \frac{a b c}{2} \int\limits_0^\infty \frac{\dd q}{(b^2 + q) f(q)}, \quad 
    L_3 = \frac{a b c}{2} \int\limits_0^\infty \frac{\dd q}{(c^2 + q) f(q)}.
\end{equation}
are the geometric factors of the corresponding axes. 
For a sphere, the geometric factors $L_1 = L_2 = L_3 = 1/3$, and Eqs.~\eqref{static_dipole} agree with the ones obtained in \cite{baresch, PRL}. For a spheroid, $L_1 = L_2$, these equations agree with Ref.~\cite{senior}.

\subsection{Radiative corrections}

Polarizabilities \eqref{static_monopole} and \eqref{static_dipole} were obtained in the quasistatic limit. This approximation does not account for the radiation from the induced multipoles. In this section we derive the radiation corrections for the polarizabilities using the energy conservation.

We assume an arbitrary monochromatic incident field, and a scattered field containing only the monopole and dipole contributions \eqref{scattered_pressure_monopole}--\eqref{scattered_velocity_dipole}. We start from the expression for the absorbed power in a monochromatic field: $P_{\text{abs}} = - \oint \limits_S {\mathbfcal{P}} \cdot \dd \vb{S}$, where $\mathbfcal{P} = \frac{1}{2}\Re(\vb{v}^* p)$ is the acoustic energy flux density, and the total fields are $(p, \vb{v}) = (p_{\text{inc}}, \vb{v}_{\text{inc}}) + (p_{\text{sc}}, \vb{v}_{\text{sc}})$, with $(p_{\text{sc}}, \vb{v}_{\text{sc}}) = (p_{\text{M}}, \vb{v}_{\text{M}}) + (p_{\text{D}}, \vb{v}_{\text{D}})$. Using the same techniques as in Appendix~\ref{app:force_derivation}, we derive
\begin{equation}
    P_{\text{abs}} = P_{\text{ext}} - P_{\text{sc}}\,,
\label{Pabs}
\end{equation}
\begin{align}
    P_{\text{ext},\mathrm{M}} &= - \frac{1}{2}  \omega  \Im \left( M^* p_{\text{inc}}\right),  & P_{\text{sc},\mathrm{M}} &= \frac{\omega \abs{M}^2 k^3}{8 \pi \beta},  \nonumber \\
    P_{\text{ext},\mathrm{D}} &= - \frac{1}{2}  \rho \omega  \Im \left( \vb{D}^* \cdot \vb{v}_{\text{inc}} \right), & P_{\text{sc},\mathrm{D}} &= \frac{\rho \omega |\mathbf{D}|^2   k^3}{24 \pi}.
    \label{eq:Pext_sc}
\end{align}
Here the extinction power $P_{\text{ext}}$ originates from the mixed terms involving both the incident and scattered fields, 
while the scattered power $P_{\text{sc}}$ involves only the scattered fields.

For a lossless particle, the absorbed power vanishes: $P_{\rm abs} = 0$. For Eqs.~\eqref{Pabs} and \eqref{eq:Pext_sc}, this leads to
\begin{equation}
\label{eq:constraint_imaginary_1}
{\Im\!\left(\alpha^{-1}_{\text{M}}\right)} = - \frac{k^3}{4 \pi} \,, \quad 
\frac{1}{2i}\left(\ten{\alpha}^{-1}_{\text{D}} - \ten{\alpha}^{-1 \, \dagger}_{\text{D}} \right) = - \ten{I} \frac{k^3}{12\pi}\,.
\end{equation}
Assuming diagonal dipole polarizability $\ten{\alpha}_{\rm D}$, Eqs.~\eqref{eq:constraint_imaginary_1} determine small imaginary radiative corrections to the real-valued quasistatic polarizabilities \eqref{static_monopole} and \eqref{static_dipole} \cite{Albaladejo2010Feb, Sipe1974PRA, LeRu2013Jan}:
\begin{align}
\label{eq:polariz_1}
({\alpha_{\text{M}}})^{-1} = ({\alpha_{\text{M}}^{\text{st}}})^{-1} - \dfrac{\iu k^3}{ 4\pi}\,,~~
\left(\ten{\alpha}_{\text{D}}\right)^{-1} = \left(\ten \alpha_{\text{D}}^{_{\text{st}}}\right)^{-1} - \dfrac{\iu k^3}{12\pi} \ten{I}\,.
\end{align}

\section{Angular stability}
\label{app:angular_dependence_torque}

\begin{figure}[h]
\centering
\includegraphics[width=0.5\linewidth]{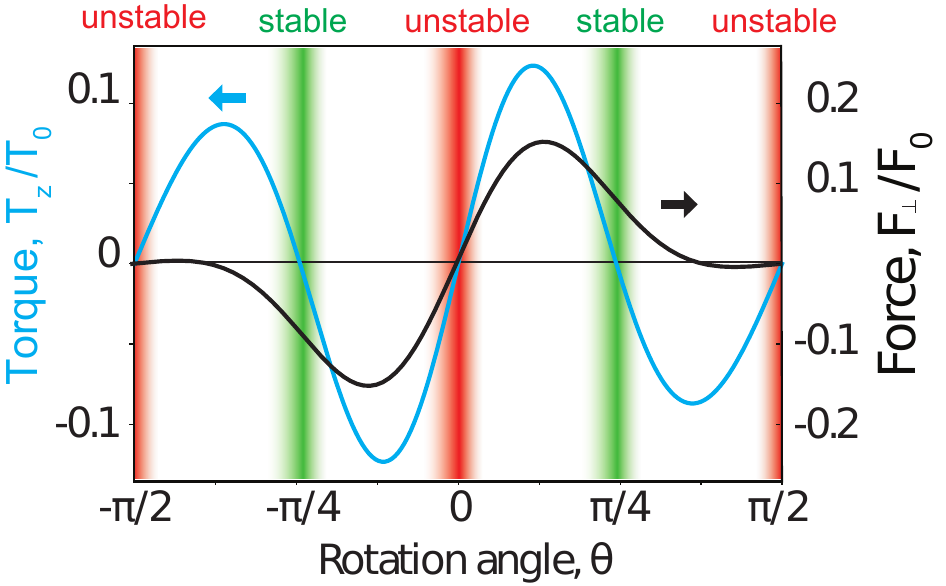}
\caption{Angular dependence of the acoustic radiation torque and force on a Mie-resonant particle considered in the main text at $ka = 3.12$. The angles of the stable and unstable torque zeroes are highlighted in green and red, respectively.}
\label{fig:stability}
\end{figure}

The angular stability of a particle is determined by conditions $T_z = 0$ and $\partial T_z / \partial \theta < 0$, which means that the particle restores its angular orientation for a small perturbation of $\theta$ near the equilibrium. 
To analyze stability of the torque zeroes in Fig.~\ref{fig:Mie_acoustomechanics}, we plot the angular dependence of the total torque computed at the frequency $ka=3.12$ (point $S_2$ in Fig.~\ref{fig:Mie_acoustomechanics}), see Fig.~\ref{fig:stability}. One can see that at this frequency, the angular stability conditions are satisfied for $\theta = \pm\pi/4$. Note that for these stable configurations the particle experiences a non-zero lateral force ${F}_\perp$. Similar analysis shows that the 
point $S_1$ in Fig.~\ref{fig:Mie_acoustomechanics}, corresponding to the frequency $ka=1.91$, is also stable for $\theta = \pm\pi/4$. In turn, the point $U$ with the frequency $ka=2.6$ does not have any stable orientations for $\theta \in (0, \pi/2)$. 

\section{Nonlocality-induced stable torque}
\label{app:nonlocality}

\begin{figure}[h]
    \centering
    \includegraphics[width=0.95\linewidth]{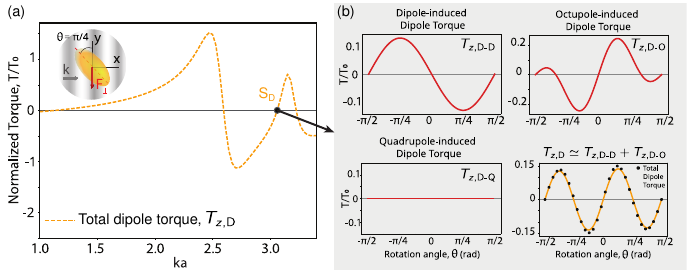}
    \caption{(a) Numerical calculations of the total dipole torque $T_{z,\text{D}}$ vs. the dimensionless parameter $ka$ for a spheroid particle from Fig.~\ref{fig:Mie_acoustomechanics} for $\theta = \pi/4$.
    (b) The dipole-dipole, dipole-quadrupole, and dipole-octupole torque contributions, as well as the total dipole torque, vs. the orientation angle $\theta$ for $ka=3.06$. The bottom right panel shows both the total dipole torque $T_{z,\text{D}}$ (symbols) and its approximation by Eq.~\eqref{eq:dipole_decomposition} (curve), justifying this approximation.   
    }
    \label{fig:nonlocality}
\end{figure}

Particles beyond the Rayleigh limit exhibit two important complications: {(i)} the higher-multipole contributions (quadrupole, octupole, etc.) to the {\it scattered} field and the torque, Fig.~\ref{fig:Mie_acoustomechanics}~(b), and {(ii)} the non-locality effects induced by the multipole contributions in the {\it incident} field, which appear even for the dipole-approximate scattered field, Fig.~\ref{fig:nonlocality}. 
In this Appendix, we analyze the nonlocality effects for the dipole torque contribution $T_{z,\text{D}}$ near the zero  $S_{\text{D}}$ in Fig.~\ref{fig:Mie_acoustomechanics}. For a spheroid dipole particle (or any particle possessing inversion symmetry), the main nonzero contributions in the expansion \eqref{eq:tz_all_multipoles} are provided by the dipole and octupole terms in the incident field:
\begin{equation}
    T_{z,\text{D}} = \sum \limits_{m=-1}^{1} T_z^{(1m)} \simeq T_{z, \text{D-D}} +  T_{z, \text{D-O}}\,.
    \label{eq:TzD_decomposition}
\end{equation}
Here, the first term, $T_{z, \text{D-D}}$, corresponds to the dipole moment $\vb{D}_{\text{D-D}} = \ten{\alpha}^{\text{D}} \vb{v}$ in Eq.~\eqref{eq:dipole_decomposition} induced by the dipole in the incident field. The second term, $T_{z, \text{D-O}}$, corresponds to the dipole moment $\vb{D}_{\text{D-O}} = \ten{\gamma}^{\text{D}} \grad \grad \vb{v}$ in Eq.~\eqref{eq:dipole_decomposition} induced by the octupole in the incident field. 
While expression \eqref{eq:torque_corr} for the dipole torque  is valid for all the dipole-moment contributions induced by the multipoles in the incident field, finding the corresponding tensors $\ten{\alpha}^{\text{D}}$ and $\ten{\gamma}^{\text{D}}$ for a non-small Mie particle is rather complicated. Instead, we use the equivalent partial-wave expansion approach to find $T_{z, \text{D-D}}$ and $T_{z, \text{D-O}}$.

Let us consider the dipole-octupole contribution $T_{z, \text{D-O}}$. It corresponds to the octupole part of the incident field, i.e., the terms $A_{3m}$, $m=-3,...,3$, in the spherical-harmonics expansion \eqref{multipolar-decomposition}. 
Solving the scattering problem for this part of the incident field, we find the dipole contribution to the scattered field, which is described by the scattering coefficients $S_{1m} = S_{1m,\text{D-O}}$ in Eq.~\eqref{multipolar-decomposition}.
Next, the dipole torque produced by this contribution is described by the $n=1$ terms in the expansion \eqref{eq:tz_all_multipoles}: $T_{z, \text{D-O}} \propto \Re \sum_{m=-1}^{1} m (A_{1m}^{*} + S_{1m,\text{D-O}}^*)S_{1m,\text{D-O}}$. Note that here the {\it dipole} part of the incident field, $A_{1m}$, is used.
Such method of calculation of the multipole moments induced by different multipoles in the incident field is closely related to the T-matrix approach (see SM of \cite{Poleva2023Jan}).

The dipole-dipole contribution to the torque, $T_{z, \text{D-D}}$, is calculated in a similar manner, involving the dipole part of the incident field, $A_{1m}$, $m=-1,...,1$, the corresponding scattering coefficients, $S_{1m} = S_{1m,\text{D-D}}$, resulting in $T_{z, \text{D-D}} \propto \Re \sum_{m=-1}^{1} m (A_{1m}^{*} + S_{1m,\text{D-D}}^*)S_{1m,\text{D-D}}$.

After calculating all necessary coefficients $A_{nm}$ and $S_{nm}$ numerically (see Materials and Methods), we plot the total dipole torque, as well as the dipole-dipole, quadrupole-dipole, and octupole-dipole contributions to it, in Fig.~\ref{fig:nonlocality}. The higher-order terms become particularly important near the torque zeros, such as $ka=3.06$ for $\theta = \pi/4$ in Fig.~\ref{fig:nonlocality}(a). The angular $\theta$-dependences of the torque contributions for $ka=3.06$ [the $S_{\text{D}}$ point in Fig.~\ref{fig:Mie_acoustomechanics}(b)] are plotted in Fig.~\ref{fig:nonlocality}(b).  
Note that the quadrupole-dipole contribution vanishes, as taken into account in Eq.~\eqref{eq:TzD_decomposition}. 
In turn, the octupole-dipole contribution $T_{z,\text{D-O}}$ is dominant and essentially determines the torque dependence on the orientation angle $\theta$. While the dipole-dipole contribution $T_{z,\text{D-D}}$ vanishes only for symmetric configurations $\theta=0,\pi/2$ (as predicted by the Rayleigh approximation analyses in Sec.~\ref{sec:Rayleigh_torque}), the octupole-dipole part $T_{z,\text{D-O}}$, as well as the total dipole torque, have two additional zeros in the intermediate range (corresponding to tilted spheroids). These additional zeros provide the necessary conditions for the stable acoustic lift.

\section{Calculation of the Acoustic Radiation Torque in 2D-axisymmetric geometry}
\label{app:numerical_axisymmetric}

\begin{figure}[b]
\centering
\includegraphics[width=0.7\linewidth]{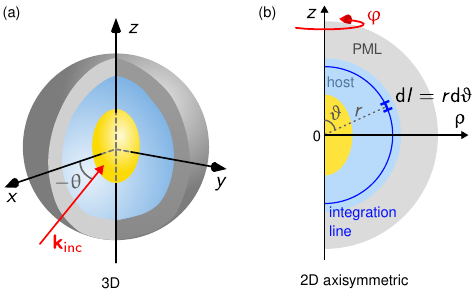}
\caption{Geometry of the 3D (a) and 2D (b) axisymmetric  problems for COMSOL simulations. The fields are calculated in the 2D model, and then can be revolved around the symmetry axis $z$. 
The outer boundary of the computational area is the perfectly matched layer (PML), introduced to absorb any outgoing waves such that there is no back reflection.}
\label{fig:2d_axisymmetric}
\end{figure}

COMSOL Multiphysics is supplied with the possibility to solve 3D acoustic scattering problems for axisymmetric objects as a series of 2D scattering problems, which significantly reduces the computational time \cite{Gladyshev2024ACSPhot}. 

The geometry of our model is shown in Fig.~\ref{fig:2d_axisymmetric}.
The incident and scattered fields can be written via the Jacobi-Anger expansion into azimuthal modes:
\begin{align}
\label{jacobi_anger}
    (p, \mathbf{v}) = \sum \limits_{m = - \infty}^{\infty} (p^m, \mathbf{v}^m) \eu^{\iu m \varphi},
\end{align}
which use the {\it cylindrical} coordinates $(\varrho, \varphi, z)$. 
The general expression for the force \eqref{force_tensor} can also use the field components in the cylindrical coordinates, but we seek for the {\it cartesian} components of the resulting force. Therefore, we write \eqref{force_tensor} in the form involving the rotation matrix 
\begin{equation}
\label{rotation_matrix}
\ten{R} = \begin{pmatrix} \cos \varphi & - \sin \varphi & 0 \\
\sin \varphi & \cos \varphi & 0 \\
0 & 0 & 1
\end{pmatrix},
\end{equation}
providing the transformation between the cylindrical and Cartesian components. This results in
\begin{align}
\label{force_cartesian_cylindrical}
F_i = - \oint\limits_S  \hat{\mathcal{T}}_{ij}  {n}_j  \dd S = - \oint\limits_S \hat{R}_{i\alpha}   \hat{\mathcal{T}}_{\alpha \beta}   \hat{R}_{\beta j}  \hat{R}_{j \beta} {n}_{\beta} \dd S = -\oint\limits_S  \hat{\mathcal{T}}_{i \beta}   {n}_\beta \dd S,
\end{align}
where the Roman and Greek indices correspond to the Cartesian and cylindrical coordinates, respectively,
${n}_{i} = \hat{R}_{i \alpha} {n}_\alpha$, $ \hat{\mathcal{T}}_{i \beta}  = \hat{R}_{i \alpha}  \hat{\mathcal{T}}_{\alpha \beta} $, and summation over repeated indices is assumed.
Explicitly,  $n_\alpha = (\varrho/ \sqrt{\varrho^2 + z^2}, 0, z / \sqrt{\varrho^2 + z^2})^{T}$ and
\begin{align}
\label{T_xyzb}
\ten{\mathcal{T}}_{i \beta} = 
\begin{pmatrix}
W \cos\varphi + v_{\varrho \varrho} \cos \phi - v_{\varphi \varrho} \sin\varphi~~ &
-W \sin \varphi + v_{\varrho \varphi} \cos \varphi - v_{\varphi \varphi} \sin \varphi~~ & 
v_{\varrho z} \cos \varphi - v_{\varphi z} \sin \varphi \\
W \sin \varphi + v_{\varphi \varrho} \cos \varphi + v_{\varrho \varrho} \sin \varphi~~ & 
W \cos \varphi + v_{\varphi \varphi} \cos \varphi + v_{\varrho \varphi} \sin \varphi~~ &
 v_{\varphi z} \cos \varphi + v_{\varrho z} \sin \varphi \\
v_{z \varrho} &
v_{z \varphi} &
 W + v_{zz} 
\end{pmatrix},  
\end{align}
where $W = \frac{1}{4} \beta_0 | p|^2  - \frac{1}{4} \rho_0  |\mathbf{v}|^2 $ and $v_{\alpha \beta} = \frac{1}{2}\rho\, {\Re}( v^*_\alpha v_\beta )$. 
Orthogonality of the cylindrical harmonics in Eq.~\eqref{jacobi_anger} allows us to reduce the surface integral in Eq.~\eqref{force_cartesian_cylindrical} to a line integral. Assuming the integration surface $S$ to be a sphere of radius $r$, we have 
\begin{equation}
    \oint \limits_{S} \dd S = \int \limits_0^{2\pi} \dd \varphi \int \limits_0^{\pi} \dd \vartheta\, r^2 \sin\vartheta = \int \limits_0^{2\pi} \dd \varphi \int \limits_0^{\pi r}  \dd l\, \varrho,
\end{equation}
where $\vartheta$ is the polar angle of spherical coordinates, $\varrho = r \sin\vartheta$, and $\dd l = r \dd \vartheta$ is the meridional line element (see Fig.~\ref{fig:2d_axisymmetric}). %
Substituting Eqs.~\eqref{jacobi_anger} and \eqref{T_xyzb} in Eq.~\eqref{force_cartesian_cylindrical}, brings about three different types of terms:
\begin{subequations}
\label{eq:vv_terms}
\begin{align}
    \frac{1}{2} \Re \int\limits_0^{2\pi} v^*_{\alpha} v_{\beta} \, \cos \varphi \: \dd \varphi \, & = \frac{ \pi}{2} \Re \left( \sum_m v_{\alpha}^m v_{\beta}^{* \, m+1} + \sum_m v_{\alpha}^{m+1} v_{\beta}^{* \, m} \right), \\
    \frac{1}{2} \Re \int \limits_0^{2\pi} v^*_{\alpha} v_{\beta} \, \sin \varphi \: \dd\varphi &= \frac{\pi}{2} \Im \left(\sum_m v_{\alpha}^m v_{\beta}^{* \, m+1} - \sum_m v_{\alpha}^{m+1} v_{\beta}^{* \, m}  \right), \\
    \frac{1}{2}\Re\int \limits_0^{2\pi} v^*_{\alpha} v_{\beta} \: \dd \varphi &= \pi  \Re  \sum_m v_{\alpha}^m v_{\beta}^{* \, m} .
\end{align}
\end{subequations}
Using these relations, 
we derive the Cartesian components of the force:
\begin{subequations}
\label{eq:F_m}
\begin{align}
    \label{eq:Fxm}
    F_x = & - \frac{\pi}{2} \sum_m \int\limits_{0}^{\pi r}  \dd l  \,\varrho  \bigg[ n_{\varrho} \Re \left( \beta p^m p^{* \, m+1} - \rho \vb{v}^m \cdot \vb{v}^{* \, m+1} + 2 \rho v_\varrho^m v_\varrho^{* \, m+1} \right) +  n_{\varrho} \rho\Im \left( v_\varrho^m v_\varphi^{* \; m+1} - v_\varrho^{m+1} v_\varphi^{* \; m} \right)
    \nonumber \\ 
    & \qquad \qquad  \qquad  \qquad + 
    n_z \rho \Re  \left(   v_\varrho^m v_z^{* \; m+1} + v_\varrho^{m+1} v_z^{* \; m} \right) +  n_z \rho \Im \left( v_\varphi^m v_z^{* \; m+1} - v_\varphi^{m+1} v_z^{* \; m} \right)\bigg],  \\ \nonumber \\
    F_z = &- \pi \Re \sum_m \int\limits_{0}^{\pi r}   \dd l \, \varrho   \left[  \frac{n_z}{2}  \left(\beta p^m p^{* \; m} - \rho \vb{v}^m \cdot \vb{v}^{* \; m} \right) + n_\varrho \rho v_\varrho^m v_z^{*\; m}   +  n_z  \rho  v_z^m v_z^{* \;m}  \right].
\end{align}
\end{subequations}
Expression for $F_y$ can be obtained by changing $\Re \to \Im$ and $\Im \to \Re$ in Eq.~\eqref{eq:Fxm}. However, the symmetry of the problem under consideration dictates $F_y = 0$.
Expressions for the torque can be obtained in a similar way:
\begin{subequations}
\label{eq:T_m}
\begin{align}
    T_y = &-\frac{\pi}{2} \sum_m \int\limits_{0}^{\pi r}  \dd l \, \varrho \bigg[  z\, n_\varrho \Re \left(\beta\, p^m p^{* \, m+1} - \rho\, v_\varrho^m v_\varrho^{* \, m+1} - \rho\, v_\varphi^m v_\varphi^{* \, m+1} - \rho\, v_z^m v_z^{* \, m+1} \right)  \nonumber
    \\& -  
    \varrho\, n_z \Re \left(\beta\, p^m p^{* \, m+1} - \rho\, v_\varrho^m v_\varrho^{* \, m+1} - \rho\, v_\varphi^m v_\varphi^{* \, m+1} - \rho\, v_z^m v_z^{* \, m+1} \right)  \nonumber
    \\&  +  
    z\, n_\varrho \Im \left( \rho\, v_\varrho^m v_\varphi^{* \, m+1} - \rho\, v_\varrho^{m+1} v_\varphi^{* \, m}\right) -  \varrho\, n_\varrho \Re \left( \rho\, v_\varrho^m v_z^{* \, m+1} + \rho\, v_\varrho^{m+1} v_z^{* \, m}\right)
    \nonumber \\& +  z\, n_\varrho \Re \left( \rho\, v_\varrho^m v_\varrho^{* \, m+1} + \rho\, v_\varrho^{m+1} v_\varrho^{* \, m}\right) +  z\, n_z \Im  \left(\rho\,
    v_\varphi^m v_z^{*\, m+1} - \rho\,
    v_\varphi^{m + 1} v_z^{*\, m} \right)  \nonumber
    \\&  + z\, n_z \Re  \left(\rho \,
    v_\varrho^m v_z^{* m+1} + \rho\,
    v_\varrho^{m + 1} v_z^{*\, m} \right) - \varrho\, n_z  \Re \left(\rho \, v_z^{m} v_z^{* \, m+1} + \rho\, v_z^{m+1} v_z^{* \, m} \right)
    \bigg],
    \label{eq:Ty_m} \\ 
    T_z =& - \pi \rho \Re \sum\limits_m \int \limits_0^{\pi r} \dd l \, \varrho  \left( n_{\varrho} v_{\varphi}^{*\, m} - n_{\varphi} v_{\varrho}^{* \, m} \right) \left( v_z^m z + v_{\varrho}^m \varrho \right),
    \label{eq:Tz_m}
\end{align}
\end{subequations}
where the mass density $\rho$ and the cylindrical coordinate $\varrho$ are not to be mixed. Due to the symmetry of the problem, $T_x = 0$.
The forces and torques \eqref{eq:F_m}--\eqref{eq:T_m} can be calculated by solving the scattering problem for a sufficient number of azimuthal modes \eqref{jacobi_anger}. 
Oblique incidence of a plane wave with the angle $-\theta$ is equivalent to the spheroid rotated by the angle $\theta$, and the force in the coordinate system of the incident wave can be restored using the corresponding rotation matrix. 
Note that the Cartesian coordinate system used in this Appendix (Fig.~\ref{fig:2d_axisymmetric}) is related to the one used in the main text (Fig.~\ref{fig:Rayleigh_acoustomechanics}) via $(F_x, F_y, F_z) \to (-F_x, F_z, F_y)$ and $(T_x, T_y, T_z) \to (-T_x, T_z, T_y)$.

\end{document}